\documentclass[]{aastex631}

\usepackage{bm}
\usepackage{amsmath}
\usepackage{graphicx}       
\usepackage{multirow}
\usepackage{soul}

\begin{document}

\title{Distinguishing Dark Matter, Modified Gravity, and Modified Inertia with the Inner and Outer Parts of Galactic Rotation Curves}

\correspondingauthor{Kyu-Hyun Chae}
\email{chae@sejong.ac.kr, kyuhyunchae@gmail.com}

\author[0000-0002-6016-2736]{Kyu-Hyun Chae}
\affiliation{Department of Physics and Astronomy, Sejong University, 209 Neungdong-ro Gwangjin-gu, Seoul 05006, Republic of Korea}



\begin{abstract}

The missing gravity in galaxies requires dark matter, or alternatively a modification of gravity or inertia. These theoretical possibilities of fundamental importance may be distinguished by the statistical relation between the observed centripetal acceleration of particles in orbital motion and the expected Newtonian acceleration from the observed distribution of baryons in galaxies. Here predictions of cold dark matter halos, modified gravity, and modified inertia are compared and tested by a statistical sample of galaxy rotation curves from the Spitzer Photometry and Accurate Rotation Curves database. Modified gravity under an estimated mean external field correctly predicts the observed statistical relation of accelerations from both the inner and outer parts of rotation curves. Taken at face value there is a $6.9\sigma$ difference between the inner and outer parts on an acceleration plane which would be inconsistent with current proposals of modified inertia. Removing galaxies with possible systematic concerns such as central bulges or special inclinations does not change this trend. Cold dark matter halos predict a systematically deviating relation from the observed one. All aspects of rotation curves are most naturally explained by modified gravity.

\end{abstract}



\section{Introduction} \label{sec:intro}
Cold (having a clumpy property on galaxy scales) dark matter (CDM) is a key ingredient of the standard $\Lambda{\rm{CDM}}$ cosmological model. CDM has been popular since the 1980s as it seemed to provide a plausible framework for modeling the large-scale structure of the universe and galaxy formation (e.g., \citealt{white1978,peebles1982,blumenthal1984}). Modern $\Lambda{\rm{CDM}}$ modeling uses hydrodynamic simulations (e.g., \citealt{dicintio2014a,tollet2016,lazar2020}) to predict the structure of the dark matter halo embedding, nurturing, and interacting with the galaxy. However, the $\Lambda{\rm{CDM}}$ model is faced with serious challenges (\citealt{famaey2012,kroupa2012,bull2016,bullock2017,divalentino2021,abdalla2022,banik2022,perivolaropoulos2022}; see also \citealt{peebles2022} for an illuminating viewpoint). The concept of DM is tied with the standard gravitational dynamics of Newton and Einstein. Standard theory is unique in that it obeys the strong equivalence principle (SEP) by which the internal dynamics of a galaxy falling freely in a constant external field from surrounding structures is unaffected by the external field. Alternative theories of gravitational dynamics that break the SEP can replace the need for DM, as was theorized in the framework called modified Newtonian dynamics \citep[MOND;][]{milgrom1983a}.

MOND makes several salient predictions \citep{milgrom1983a,milgrom1983b} about galactic kinematics including the baryonic Tully-Fisher relation, the radial acceleration relation (RAR), and the external field effect (EFE) which is a manifestation of the SEP's breakdown.  All these predictions were confirmed or being tested favorably by current data (e.g., \citealt{mcgaugh2005,mcgaugh2016,chae2020b,chae2021,chae2022b}). Theoretically, MOND can be realized by modified gravity \citep{bekenstein1984,milgrom2010}, i.e., modified Poisson equations, or by modified inertia \citep{milgrom1994,milgrom1999,milgrom2022}, i.e., a modification of Newton’s laws of motion, at accelerations weaker than about ${10^{ - 10}}~{\rm{m }}~{{\rm{s}}^{ - 2}}$.

A rotation curve (RC) covering a large dynamic range from the higher-acceleration inner rising part to the lower-acceleration far outskirts may have an interesting discriminating power in testing CDM, modified gravity, and modified inertia. The observed circular rotation speed $V$ at radius $R$ in the midplane of the disk in a rotationally supported galaxy implies a radial (i.e., centripetal) acceleration
\begin{equation}
    {g_{{\rm{obs}}}}(R) = \frac{{{V^2}}}{R},
    \label{eq:gobs}
\end{equation}
while the observed baryonic (i.e., stars and gas) mass distribution ${\rho _{{\rm{bar}}}}(R,z)$ implies a Newtonian radial acceleration 
\begin{equation}
    {g_{{\rm{bar}}}}(R) = {\left| {\frac{{\partial {\Phi _{\rm{N}}}(R,z)}}{{\partial R}}} \right|_{z = 0}},
    \label{eq:gbar}
\end{equation}
where ${\Phi _{\rm{N}}}(R,z)$ is the Newtonian potential satisfying the Poisson equation
\begin{equation}
	{\nabla ^2}{\Phi _{\rm{N}}} = 4\pi G{\rho _{{\rm{bar}}}},
    \label{eq:poisson}
\end{equation}
where $G$ is Newton’s gravitational constant. The empirical relation between ${g_{{\rm{obs}}}}$ and ${g_{{\rm{bar}}}}$ (the RAR) provides an interesting testbed for competing theories. 

This work considers 152 galaxies with good-quality RCs selected from the Spitzer Photometry and Accurate Rotation Curves (SPARC) database \citep{lelli2016}. Recently, \cite{chae2022b} tested MOND-modified gravity theories (AQUAL and QUMOND) with the outer part of RCs. They find that AQUAL is preferred over QUMOND because the AQUAL-required external field strength is well consistent with the value expected from cosmic environments while the QUMOND-required value is a little higher than the expected value. It is then interesting to ask whether AQUAL can predict correctly the observed inner part of RCs. Unlike the outer RCs, numerically predicted properties of the inner RCs under AQUAL (also QUMOND) are so complex that they cannot be described by a single model curve on a acceleration plane \citep{chae2022a}. As \cite{chae2022a} demonstrated for various configurations, AQUAL and QUMOND unambiguously predict that the inner part RCs deviate, though by a small amount, from the algebraic MOND relation even when the inner part is in a supercritical acceleration regime (see, also, \citealt{brada1995,angus2012}). However, modified inertia does not predict such deviations in the inner RCs \citep{milgrom1994,milgrom2022}.

Thus, the statistical distribution of $(g_{{\rm{bar}}},g_{{\rm{obs}}})$ on the acceleration plane can be used to distinguish between modified gravity and modified inertia. The selected 152 galaxies provide 3097 data points. These data points can be considered virtually independent (see \citealt{mcgaugh2016}) because $g_{\rm{obs}}$ is measured from an independent ring and $g_{{\rm{bar}}}$ is derived at each ring radius from the observed baryonic mass distribution. In this work the AQUAL field equation is numerically solved for each SPARC galaxy and the simulated/predicted distribution of $(g_{{\rm{bar}}},g_{{\rm{obs}}})$ is compared with the observed data points on the acceleration plane. For a subsample of 65 galaxies that have well-defined outer RCs, Bayesian inferred/predicted  $(g_{{\rm{bar}}},g_{{\rm{obs}}})$ under the AQUAL theory \citep{chae2022b} are also compared to the SPARC data points. 

This work also performs Bayesian modeling of the 152 SPARC RCs with DM halos predicted by $\Lambda{\rm{CDM}}$ hydrodynamic simulations. The posterior distribution of $(g_{{\rm{bar}}},g_{{\rm{obs}}})$ from the DM halo modeling results is compared with the SPARC data points on the acceleration plane. Also, for the subsample of 65 galaxies with well-defined outer RCs the DM halo models are tested against the AQUAL model through a Bayesian statistic.

For the first time, dark matter, modified gravity, and modified inertia are simultaneously tested and distinguished by considering the inner and outer parts of galactic rotation curves, together and separately. The paper is organized as follows. In Section~\ref{sec:data}, the data used in this work are described with an emphasis on the inner and outer parts of RCs. Section~\ref{sec:method} describes the methods concisely. Section~\ref{sec:results} describes the main results of this work. Section~\ref{sec:discussion} discusses the main results and draws a conclusion. The details of the methods can be found in Appendix~\ref{sec:appA}. The properties of DM halos derived from halo modeling are described in Appendix~\ref{sec:appB}.

\section{Data and Methods}  \label{sec:data_method}

\subsection{Data}  \label{sec:data}

The SPARC database \citep{lelli2016} provides RCs and mass distributions of stars and gas for 175 rotationally supported galaxies ranging from dwarf galaxies to giant spiral galaxies. Rotation velocities ${V_{{\rm{obs}}}}(R)$ as a function of cylindrical radius $R$ are reported for a measured inclination angle ${i_{{\rm{obs}}}}$. Baryonic mass distributions in the stellar and gas disks are reported as Newtonian rotation velocities ${V_{{\rm{disk}}}}(R)$ and ${V_{{\rm{gas}}}}(R)$ via Poisson’s equation. Similarly, the stellar mass distribution in a bulge is reported as ${V_{{\rm{bulge}}}}(R)$. The SPARC-reported values of ${V_{{\rm{disk}}}}(R)$ and ${V_{{\rm{bulge}}}}(R)$ are based on stellar mass-to-light ratios at a wavelength of $3.6\mu {\rm{m}}$ of ${\Upsilon _{{\rm{disk}}}} = 0.5{\Upsilon _ \star }$ and ${\Upsilon _{{\rm{bulge}}}} = 0.7{\Upsilon _ \star }$, where ${\Upsilon _ \star }$ is the solar value. The SPARC-reported values of ${V_{{\rm{gas}}}}(R)$ are for a gas-to-HI (neutral hydrogen) mass ratio of 1.33. The Newtonian rotation velocity of a test particle due to all observed baryons is given as ${V_{{\rm{bar}}}} = \sqrt {V_{{\rm{disk}}}^2 + V_{{\rm{bulge}}}^2 + V_{{\rm{gas}}}^2} $.

\begin{figure}
  \centering
  \includegraphics[width=1.\linewidth]{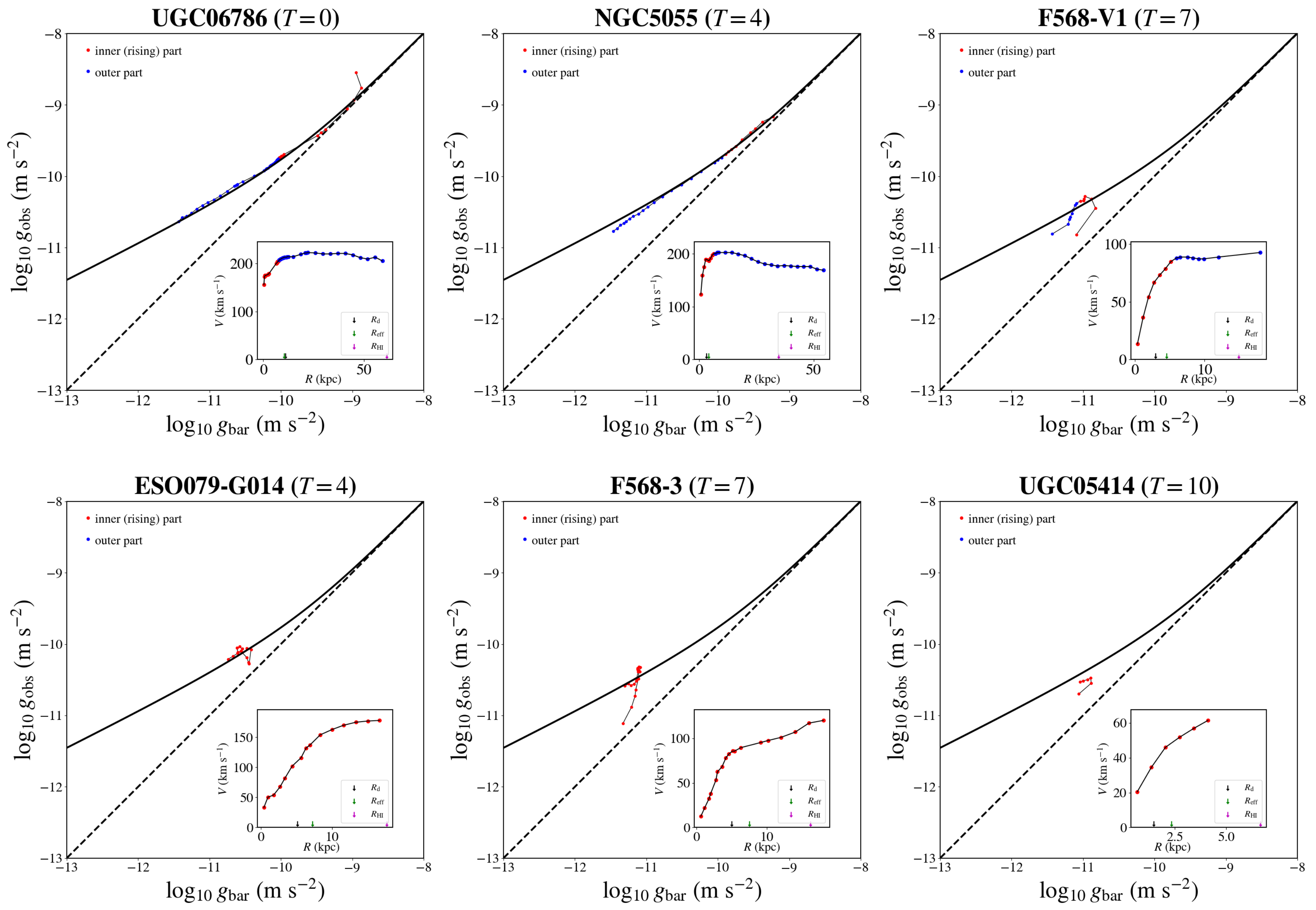}
    \vspace{-0.5truecm}
    \caption{\small 
    Examples of separating a rotation curve into the inner and outer parts. Upper row: three rotation curves of different galaxy type $T$ (as given in \citealt{lelli2016}) are shown in the acceleration space based on AQUAL-fitted parameters \citep{chae2022b}. The inset shows the SPARC-reported rotation curves. The inner rising and outer quasi-flat parts were separated visually. A clear inner-outer separation such as these can be done for about 70\% of SPARC galaxies (see the text). Lower row: examples in which the outer quasi-flat part is unobserved or ill defined from the SPARC database. For the galaxies without the outer part, AQUAL-fitting \citep{chae2022b} is not available and thus original SPARC mass models are shown here.
    } 
   \label{examples}
\end{figure} 

RCs are classified by a quality flag $Q$. There are 163 galaxies with RCs of good ($Q = 1$) or acceptable ($Q = 2$) quality. However, a recent study \citep{chae2021} finds that the RC of one galaxy (UGC 06787) is problematic, so this galaxy is excluded. In this study, the RAR based on the SPARC-reported ${V_{{\rm{obs}}}}(R)$ will be used to test various modeling and simulation results. Hence, nearly face-on galaxies with ${i_{{\rm{obs}}}} < {30^\circ }$ are excluded as in \cite{mcgaugh2016} and \cite{chae2020b} to minimize $1/\sin {i_{{\rm{obs}}}}$ corrections in ${V_{{\rm{obs}}}}(R)$.\footnote{For a recent cautionary tale in this regard, see \cite{banik2022a}.} Then, we are left with 152 galaxies. These galaxies provide 3097 independent values of ${V_{{\rm{obs}}}}(R)$ and the same number of ${V_{{\rm{bar}}}}(R)$. Previous studies \citep{mcgaugh2016,chae2020b} applied a tight signal-to-noise ratio (S/N) cut on individual rotation velocities of ${V_{{\rm{obs}}}}(R)$: velocities with ${\rm{S/N}} \le 10$ were not used. Such a cut would have discarded 469 velocities from the total of 3097 velocities from the chosen 152 galaxies. In this study, this cut is not applied for two reasons. First, the median of residuals from the RAR in a bin will be calculated though a Monte Carlo method fully considering individual errors. Thus, data with larger errors are proportionately less weighted, so that an artificial hard cut is not needed.  Second and more importantly, velocities with ${\rm{S/N}} \le 10$ are mostly from the inner rising part of the RCs. Because the inner rising part can be used to distinguish between modified gravity and modified inertia, it is better to keep all velocities of the inner part to maximize the statistical power.

The inner and outer parts of an RC are determined visually because observed RCs have various shapes. Examples of the separation of a rotation curve into the inner and outer parts can be found in the upper row of Figure~\ref{examples}. There are also extreme cases where the RC rises up to the last measured point as shown in the lower row of Figure~\ref{examples}, or no rising part is observed at all. A high-density bulge or a baryon-dominated extended gas disk can play a significant role. Thus, the transition point between the inner rising part and the outer quasi-flat part can vary greatly. For about 30\% of galaxies, the RC rises up to the last measured point (although it is not so clear-cut in some cases) so that the outer part is absent or ill defined. In those cases, the outermost measured radius is simply taken as (the lower limit of) the transition radius $R_\text{turn}$ and thus all data points belong to the inner part.\footnote{While a lower limit can only be set on the transition radius in such cases, the membership of data points in the inner part is unchanged by the underestimated transition radius. As long as the membership is correct, this work is unaffected.} The transition radius ranges from $ \approx 0$ to $ \approx 9{R_{\rm{d}}}$ where ${R_{\rm{d}}}$ is the stellar disk scale radius (see Appendix~\ref{sec:appA}). 

\begin{figure}
  \centering
  \includegraphics[width=0.7\linewidth]{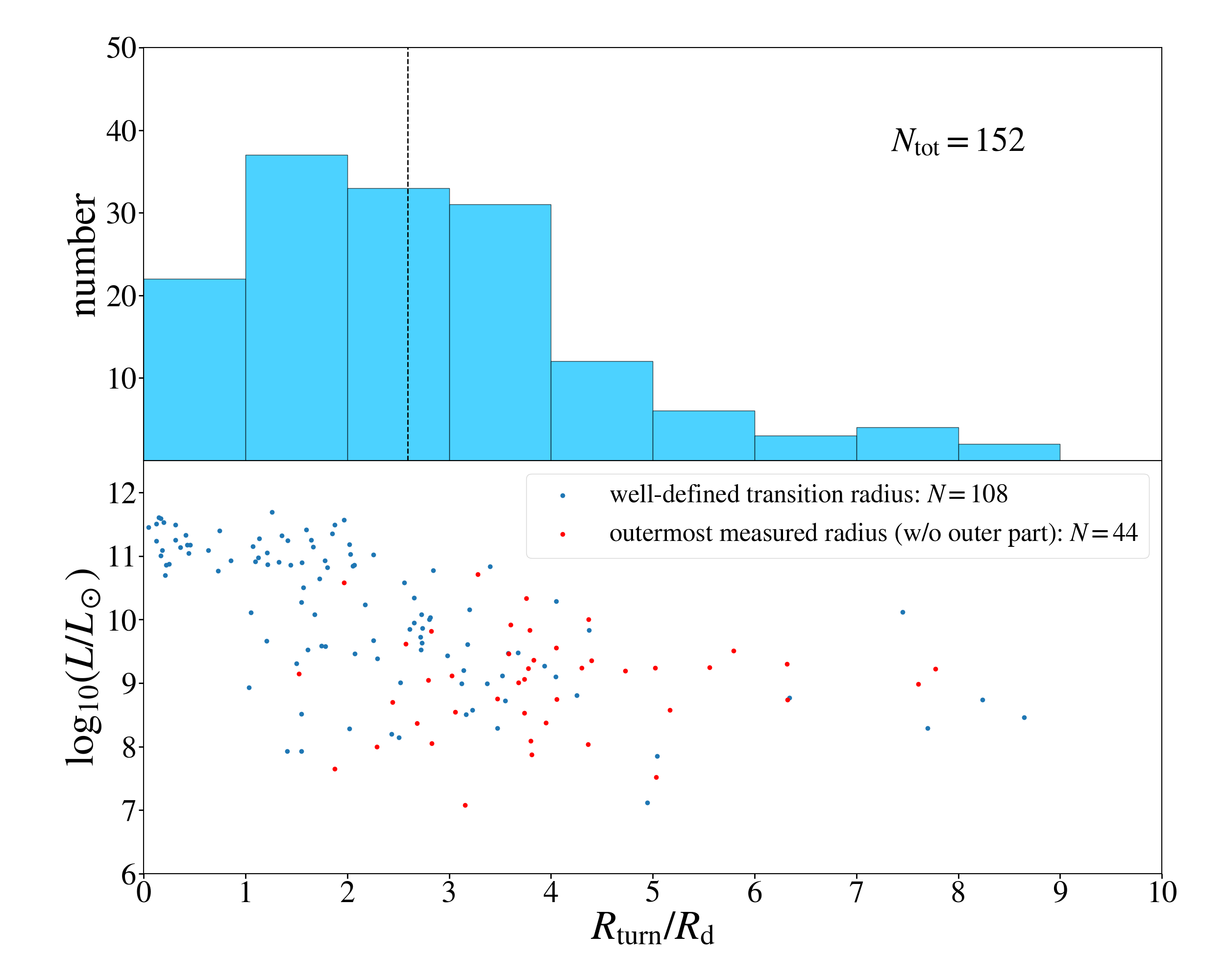}
    \vspace{-0.2truecm}
    \caption{\small 
    The upper panel shows the distribution of the inner-outer transition radius $R_\text{turn}$ for 152 SPARC rotation curves used in this study. The vertical dotted line indicates the median value of $2.6{R_{\rm{d}}}$ where ${R_{\rm{d}}}$ is the disk scale radius. The lower panel shows an anticorrelation of $R_\text{turn}/R_\text{d}$ with galaxy infrared luminosity. Red dots indicate the outermost radii when the observed RC rises up to the last measured point so that the outer part is unobserved or ill defined.
    } 
   \label{Rturn}
\end{figure}

The distribution of visually determined transition radii for the selected 152 galaxies can be found in Figure~\ref{Rturn}. The median transition radius is $2.6{R_{\rm{d}}}$. The ratio $R_\text{turn}/R_\text{d}$ is anticorrelated with galaxy luminosity as shown in the lower panel of Figure~\ref{Rturn}. This is consistent with the empirical fact that giant spiral galaxies have an extended outer part and a relatively small inner part while subluminous galaxies often have an RC rising over a large radial range. These trends are very similar if all 175 SPARC galaxies are used.

\subsection{Methods}  \label{sec:method}

To test and compare the theoretical possibilities of DM and MOND, the predicted statistical distributions of $(g_\text{bar},g_\text{obs})$ are calculated and compared with the SPARC-reported data points. For $\Lambda$CDM, a Bayesian approach is used to fit each RC of the selected 152 galaxies with $\Lambda$CDM predicted DM halos. The Bayesian inference of the model parameters is obtained through a Markov chain Monte Carlo \citep[MCMC;][]{emcee} method. The methodology is similar to that used in \cite{chae2020b}, \cite{chae2022b} and is described in detail in the Appendix~\ref{sec:appA}. The Bayesian modeling returns a predicted statistical distribution of $(g_\text{bar},g_\text{obs})$ for the inner and outer parts of the RCs. As a by-product, the estimated Bayesian information criterion \citep[BIC;][]{kass1995} is recorded for each model of each galaxy. Both the statistical/stacked distribution of $(g_\text{bar},g_\text{obs})$ from all galaxies and the individual BIC values will be used to test and distinguish models.

For modified gravity, the AQUAL field equation is used to predict $(g_\text{bar},g_\text{obs})$. Two approaches are used. For a subsample of 65 galaxies with well-defined outer RCs having wide dynamic ranges, the Bayesian modeling results from \cite{chae2022b} are used to obtain a statistical distribution of $(g_\text{bar},g_\text{obs})$. The BIC values are also estimated for these galaxies. Since the Bayesian approach is not possible for the rest of the galaxies, another approach is also considered. Based on the SPARC-reported galaxy parameters, the AQUAL field equation is numerically solved for each of all 152 galaxies using the \cite{chae2022a} AQUAL numerical solver. Observed uncertainties are applied to the numerical solutions to obtain simulated $(g_\text{bar},g_\text{obs})$.

For modified inertia, no modeling or simulation is performed. Modified inertia is simply tested by comparing the SPARC data points of the inner parts with those of the outer parts on the acceleration plane. For this, the Bayesian approach suggested by \cite{chae2022b} is used to fit the data points with a model curve.

\subsubsection{$\Lambda$CDM}  \label{sec:method_LCDM}

A widely used functional form \citep{zhao1996} to describe the DM halo is given by 
\begin{equation}
	{\rho _{{\rm{halo}}}}(r) = \frac{{{\rho _{\rm{s}}}}}{{{{\left( {\frac{r}{{{r_{\rm{s}}}}}} \right)}^\gamma }{{\left[ {1 + {{\left( {\frac{r}{{{r_{\rm{s}}}}}} \right)}^\alpha }} \right]}^{(\beta  - \gamma )/\alpha }}}},
    \label{eq:zhao}
\end{equation}
where ${r_{\rm{s}}}$ is the scale radius and ${\rho_{\rm{s}}}$ is the scale density. The virial radius ${r_{200}}$ of the halo is defined to be the radius within which the CDM density is equal to 200 times the critical density of the universe, while the concentration of the halo is defined to be ${c_{200}} \equiv {r_{200}}/{r_s}$. The halo virial mass ${M_{200}}$ is defined to be the mass within ${r_{200}}$. Parameters $\gamma$ and $\beta$ control the inner and outer density slopes while parameter $\alpha $ controls the sharpness of the inner-outer transition. The parameter choice $(\alpha ,\beta ,\gamma ) = (1,3,1)$ is the Navarro-Frenk-White (NFW) profile \citep{NFW} which is the prediction of $N$-body simulations of CDM without any baryonic physics.

Recent cosmological hydrodynamic simulations of galaxy formation within CDM halos predict diverse (stellar-to-halo mass ratio ${M_ \star }/{M_{200}}$-dependent) DM density profiles \citep{dicintio2014b,dekel2017,lazar2020} modified from the NFW profile due to baryonic feedback (e.g.\ \citealt{gnedin2002,pontzen2012}). These results provide constraints on $(\alpha ,\beta ,\gamma )$. Two specific cases are considered here. 

One is the choice $(\alpha ,\beta ) = (0.5,3.5)$ preferred by \cite{dekel2017} based on the Numerical Investigation of Hundred Astrophysical Objects (NIHAO) simulations. In this case there are three free parameters which are chosen to be ${M_{200}}$, ${c_{200}}$ and $\gamma $. (Note that ${M_{200}}$ and ${c_{200}}$ are used instead of ${r_{\rm{s}}}$ and ${\rho_{\rm{s}}}$.)  In the Bayesian inference, two prior constraints are imposed: the popular halo mass$-$stellar mass abundance matching relation \citep{behroozi2013,kravtsov2018} and the ${M_\star }/{M_{200}}$-dependent slope estimated between radii $0.01{r_{200}}$ and $0.02{r_{200}}$ from the NIHAO simulations \citep{tollet2016}. 

The other case, dubbed as the DC14 model \citep{dicintio2014b}, provides constraints on all three indices $(\alpha ,\beta ,\gamma )$ and the concentration ${c_{200}}$. In this case, all five halo parameters ${M_{200}}$, ${c_{200}}$, $\alpha $, $\beta $, and $\gamma $ are allowed to vary with five prior constraints (including the ${M_{200}}$-${M_\star }$ abundance matching relation) imposed on them. 

Another functional form of DM halos considered here is the “core-Einasto profile” proposed recently based on the Feedback In Realistic Environments (FIRE)-2 galaxy formation simulations \citep{lazar2020}. The FIRE-2 simulations provide a constraint on the core size as a function of the stellar-to-halo mass ratio. See Appendix~\ref{sec:appA} for the details that include a table summarizing the parameters and priors for each model.

\subsubsection{MOND}  \label{sec:method_MOND}

For modified gravity, the AQUAL theory field equation 
\begin{equation}
	\nabla  \cdot \left[ {\mu (\left| {\nabla \Phi } \right|/{a_0})\nabla \Phi } \right] = 4\pi G{\rho _{{\rm{bar}}}}	
    \label{eq:aqual}
\end{equation}
is numerically solved for the MOND potential $\Phi$ from which follows the radial acceleration ${g_R} = \left| {\partial \Phi /\partial R} \right|$. Here ${a_0}$ is the MOND critical acceleration measured to be $ \simeq 1.2 \times {10^{ - 10}}{\rm{m }}{{\rm{s}}^{ - 2}}$ \citep{mcgaugh2016} and $\mu (x)$ is the MOND interpolating function (IF) which is assumed to take the simple form $x/(1 + x)$ \citep{famaey2005}.\footnote{See \cite{chae2019} for why this IF is a good approximation for the acceleration range of RCs.} A numerical solution of Equation~(\ref{eq:aqual}) for the quasi-flat outer RC of a flattened system under a constant external field is given as \citep{chae2022a}
\begin{equation}
	{g_{{\rm{MOND}}}} = {g_{{\rm{bar}}}}\nu ({y_{1.1}})\left[ {1 + {\rm{tanh}}{{\left( {\frac{{1.1{e_{\rm{N}}}}}{{{g_{{\rm{bar}}}}/{a_0}}}} \right)}^{1.2}}\frac{{\hat \nu ({y_{1.1}})}}{3}} \right],
    \label{eq:aqformula}
\end{equation}
where ${y_{1.1}} \equiv \sqrt {{{({g_{{\rm{bar}}}}/{a_0})}^2} + {{(1.1{e_{\rm{N}}})}^2}} $, $\nu (y) = 0.5 + \sqrt {0.25 + {y^{ - 1}}} $, and  $\widehat \nu (y) \equiv d{\rm{ln}}\nu {\rm{(}}y{\rm{)/}}d\ln y$. This equation has the right asymptotic limits if the external field dominates over the internal field (see equation~(23) of \citealt{zonoozi2021}).

Equation~(\ref{eq:aqformula}) was recently fitted to the stacked data from the outer rotation curves by \cite{chae2022b}.  From this fit, the mean Newtonian external field strength was estimated to be  ${g_{{\rm{N,ext}}}} \approx 6.1 \times {10^{ - 3}}{a_0}$ or $\tilde e \equiv \sqrt{e_{\rm{N}}} \equiv \sqrt{g_{\rm{N,ext}}/a_0}  = 0.078_{- 0.013}^{ + 0.010}$ \citep{chae2022b}. $\tilde{e}$ is equivalent to the MOND external field in units of ${a_0}$ to a good approximation because $\tilde{e} \ll 1$. Then, under the mean value $\tilde e = 0.078$ \citep{chae2022b}, Equation~(\ref{eq:aqual}) is solved using the \cite{chae2022a} code for ${\rho _{{\rm{bar}}}}$ of each galaxy with a stellar disk and a gas disk (and a bulge if present) to predict a RAR curve for each galaxy. At each SPARC-reported radius along the curve, the actual SPARC-reported uncertainties are applied in a Monte Carlo way to obtain simulated $(g_\text{bar},g_\text{obs})$. Further details can be found in Appendix~\ref{sec:appA}.8. 

In addition to the above simulated RARs, Bayesian inferred RARs are considered for a subset of 65 galaxies whose outer rotation curves have large dynamic ranges \citep{chae2022b}. 

\section{Results}  \label{sec:results}

\subsection{Dark Matter} \label{sec:results_DM}

\begin{figure}
  \centering
  \includegraphics[width=1.\linewidth]{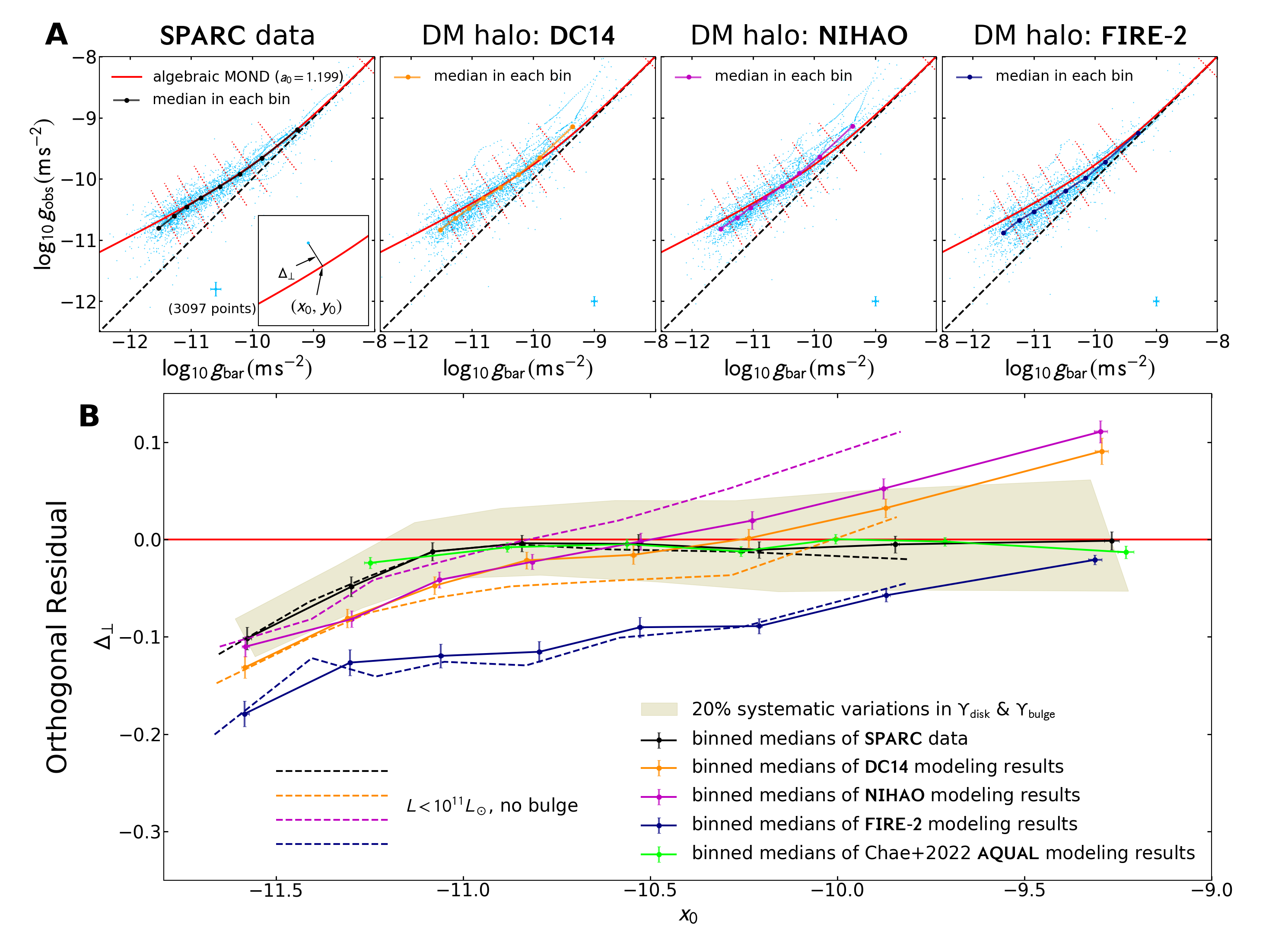}
    \vspace{-0.5truecm}
    \caption{\small 
    Comparison of Bayesian fit results for two DM halo models with SPARC data. (A) The leftmost panel shows the SPARC data of 3097 points from 152 galaxies with mass-to-light ratios ${\Upsilon _{{\rm{disk}}}} = 0.5{\Upsilon _ \odot }$ for the disk and  ${\Upsilon _{{\rm{bulge}}}} = 0.7{\Upsilon _ \odot }$ for the bulge. Typical uncertainties of the logarithm of accelerations are indicated by errorbars at the bottom of the panel.  The inset defines the orthogonal residual ${\Delta _ \bot }$from the algebraic MOND relation, i.e., the interpolating function, for ${a_0} = 1.199 \times {10^{ - 10}}{\rm{m }}{{\rm{s}}^{ - 2}}$ fitted to the outer rotation curves. The other panels show MCMC fit results for the DC14 \citep{dicintio2014a,dicintio2014b}, NIHAO \citep{tollet2016,dekel2017} and FIRE-2 \citep{lazar2020} models for hypothetical CDM halos. (B) This panel shows the trend of the medians of orthogonal residuals in the orthogonal bins. DM halo models systematically deviate from the SPARC data, in contrast to the AQUAL model \citep{chae2022b}. Dashed lines are results for a subsample of 111 galaxies without a bulge and with luminosity $L < {10^{11}}{L_ \odot }$, which reduces the role of BH feedback.
    } 
   \label{RAR_DM}
\end{figure}

Figure~\ref{RAR_DM} shows the MCMC fit results for three DM halo models and compares them with the SPARC data assuming mass-to-light ratios at a wavelength of $3.6\mu {\rm{m}}$ of ${\Upsilon _{{\rm{disk}}}} = 0.5{\Upsilon _ \odot }$ for the disk, and ${\Upsilon _{{\rm{bulge}}}} = 0.7{\Upsilon _ \odot }$ for the bulge if present. The DM halo fitted results are roughly similar to the SPARC RAR but have larger scatters (see below for further details). Relatively speaking, the DC14 model matches the SPARC RAR better than the other halo models. Figure~\ref{RAR_DM}(b) further shows the binned medians of orthogonal residuals from the algebraic MOND relation (i.e., the interpolating function represented by the red curve). The medians are estimated through a Monte Carlo method using individual Gaussian widths for ${\rm{lo}}{{\rm{g}}_{10}}{g_{{\rm{bar}}}}$ and ${\rm{lo}}{{\rm{g}}_{10}}{g_{{\rm{obs}}}}$ that incorporate all uncertainties including mass-to-light ratios. The trend of medians deviates systematically from the SPARC median in all three cases. In particular, in the high and/or low acceleration limits the models deviate outside the expected range estimated conservatively based on 20\% systematic variations in mass-to-light ratios. In fact, some galaxy-to-galaxy scatter in ${\Upsilon _{{\rm{disk}}}}$ and ${\Upsilon _{{\rm{bulge}}}}$ is expected, but the uncertainty of the median is less than 5\% \citep{schombert2022}. 

Considering that hydrodynamic simulations may be less accurate for massive galaxies due to central black holes \citep{maccio2020} and reported rotation velocities can be less accurate in the central region if a bulge is present, we also consider a subsample of 111 galaxies with luminosities $L < {10^{11}}{L_ \odot }$ and without a reported bulge. The results are represented by dashed lines in Figure~\ref{RAR_DM}(b). The results are somewhat shifted for some acceleration ranges of the DC14 and NIHAO models, but the shifted results are still discrepant with the SPARC data. Moreover, as will be shown below, if a rotation curve is separated into the inner rising and the outer quasi-flat parts, it is even harder to match the separated RARs by these DM models. 

Figure~\ref{RAR_DM}(b) also shows Bayesian inferred results with the AQUAL model for the subsample of 65 galaxies with good dynamics range outer RCs from \cite{chae2022b}. Unlike the DM halo modeling results, the AQUAL-predicted RAR matches remarkably well the SPARC RAR. These tests by the orthogonal residual $\Delta_\bot$ indicate that widely used criteria in statistical testing such as BIC may also show a positive evidence for the AQUAL model over the DM halo models from the individual modeling results of galaxies.

BIC is defined as usual by $\text{BIC}\equiv -2\ln \mathcal{L}_\text{max} + k_\text{prmt} \ln N_\text{RC-points}$ where $\mathcal{L}_\text{max}$ is the maximized value of the likelihood defined in Appendix~\ref{sec:appA}, $k_\text{prmt}$ is the number of the varied parameters of the given model, and $N_\text{RC-points}$ is the number of the fitted data points of the given RC. For 63 galaxies with inclination $>30^\circ$ out of the 65 galaxies with good dynamic range outer RCs, $\Delta\text{BIC}$ is calculated between the AQUAL model and the DC14 halo model. The latter is the overall preferred model among the tested halo models based on $\Delta_\bot$.

For the AQUAL model, $k_\text{prmt}=7(6)$ with (without) a bulge if $a_0$ is varied. But, varying $a_0$ improves little $\mathcal{L}_\text{max}$. Thus, for a fair test the AQUAL BIC is defined for the recalculated models with a fixed $a_0=1.20$. For the DC14 model, $k_\text{prmt}=10(9)$ with (without) a bulge if $\alpha$ and $\beta$ in Equation~(\ref{eq:zhao}) are varied along with $\gamma$. But, varying $\alpha$ and $\beta$ improves little ${\cal L}_\text{max}$. Thus, for a fair test, the DC14 BIC is defined for the recalculated models with fixed $\alpha$ and $\beta$ as given in Appendix~\ref{sec:appA}.6. The left panel of Figure~\ref{BIC} shows $\Delta\text{BIC}=\text{BIC}_\text{DC14}-\text{BIC}_\text{AQUAL}$ for the 63 galaxies. For the DC14 model, two cases are considered. In one case, only the outer RC is fitted as for the AQUAL modeling. In the other case, the entire RC including the inner RC is fitted as the halo model is intended to describe all radial ranges. In the latter case, BIC is calculated using the outer RC only for a fair comparison with the AQUAL case. Usually, $2<\Delta\text{BIC}<6$ is interpreted as a positive evidence, $6<\Delta\text{BIC}<10$ a strong evidence, and $\Delta\text{BIC}>10$ a very strong evidence. There are individual cases where either AQUAL or DC14 is preferred to a varying degree. This may indicate uncertainties of uncertainties in the data. However, in terms of the median $\Delta$BIC, there is some positive evidence for the AQUAL model by the usual criterion, but not a strong evidence.

\begin{figure}
  \centering
  \includegraphics[width=1.\linewidth]{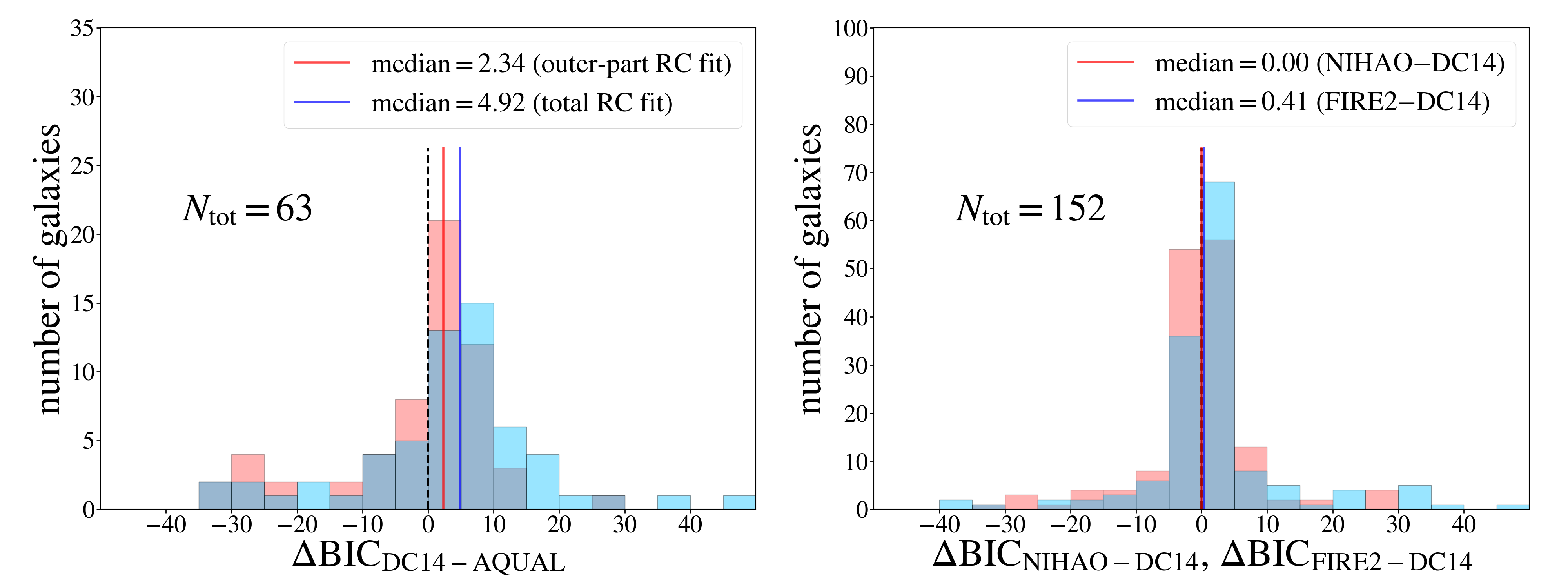}
    \vspace{-0.3truecm}
    \caption{\small 
    Comparison of models based on the Bayesian information criterion (BIC). (Left) Comparison between the DC14 halo model and the AQUAL modified gravity model. Only 63 galaxies with inclination $>30^\circ$ and outer RCs with good dynamic ranges are shown because the AQUAL modeling was performed with the outer RCs \citep{chae2022b}. For the DC14 model, two cases are considered: fitting the outer RC only, and fitting the entire RC (see the text for details). There are considerable scatters individually. However, from the median there is a positive evidence for the AQUAL model consistent with the test by $\Delta_\bot$ in Figure~\ref{RAR_DM}(b). (Right) Comparing the NIHAO and the FIRE-2 halo models with the DC14 model. The NIHAO model is statistically indistinguishable from the DC14 model. Interestingly, the FIRE-2 model appears not worse than the DC14 model unlike the test based on $\Delta_\bot$.
    }
   \label{BIC}
\end{figure} 

The right panel of Figure~\ref{BIC} shows $\Delta\text{BIC}$ for the NIHAO and FIRE-2 models with respect to the DC14 model from the modeling results of all 152 galaxies based on the entire RCs. As expected from Figure~\ref{RAR_DM}, the NIHAO model is statistically indistinguishable from the DC14 model. Interestingly, the FIRE-2 model is not significantly worse than the DC14 model based on $\Delta$BIC alone although $\Delta_\bot$ shows a significant difference between the two models (Figure~\ref{RAR_DM}(b)).

\begin{figure}
  \centering
  \includegraphics[width=1.\linewidth]{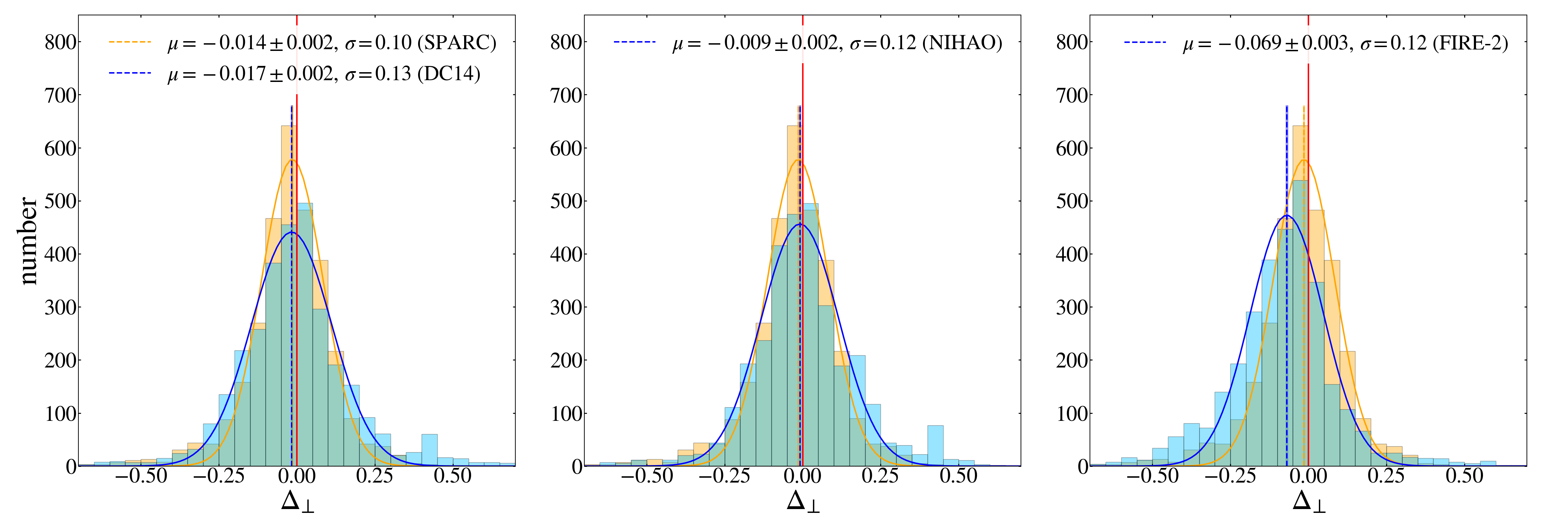}
    \vspace{-0.3truecm}
    \caption{\small 
    The distribution of orthogonal residuals $\Delta_\bot$ predicted by the DM halo models, as displayed in Figure~\ref{RAR_DM}, is compared with that of the SPARC data. From left to right: the DC14 model, the NIHAO model, and the FIRE-2 model in comparison to the SPARC data. The SPARC distribution is represented by the orange histogram while each DM halo predicted distribution is represented by the blue histogram. The DC14 model predicts a 30\% larger scatter than the SPARC data while the overall median agrees with the SPARC median. The NIHAO and FIRE-2 models have a 20\% larger scatter but the overall medians are shifted from the SPARC median. 
    } 
   \label{histodist}
\end{figure}

Finally, it is interesting to examine the overall scatters in the acceleration plane predicted by the DM halo models, independently of the running of the medians shown in Figure~\ref{RAR_DM}. Figure~\ref{histodist} shows the distributions of orthogonal residuals. The DC14 model predicts the overall median consistent with the SPARC median (though the running with acceleration is systematically deviating as shown in Figure~\ref{RAR_DM}). However, the NIHAO model predicts a somewhat shifted median and the FIRE-2 model predicts an acceptably shifted median. All three models predict 20\% to 30\% larger scatters than the SPARC scatter. In contrast, the AQUAL model (or an algebraic MOND model as investigated in detail in \citealt{li2018}) predicts a smaller scatter than the SPARC scatter, although it provides overall a better fit to the RCs than the DM halo models according to $\Delta$BIC. 

\subsection{Modified Gravity and Modified Inertia}  \label{sec:results_MOND}

\begin{figure}
  \centering
  \includegraphics[width=0.9\linewidth]{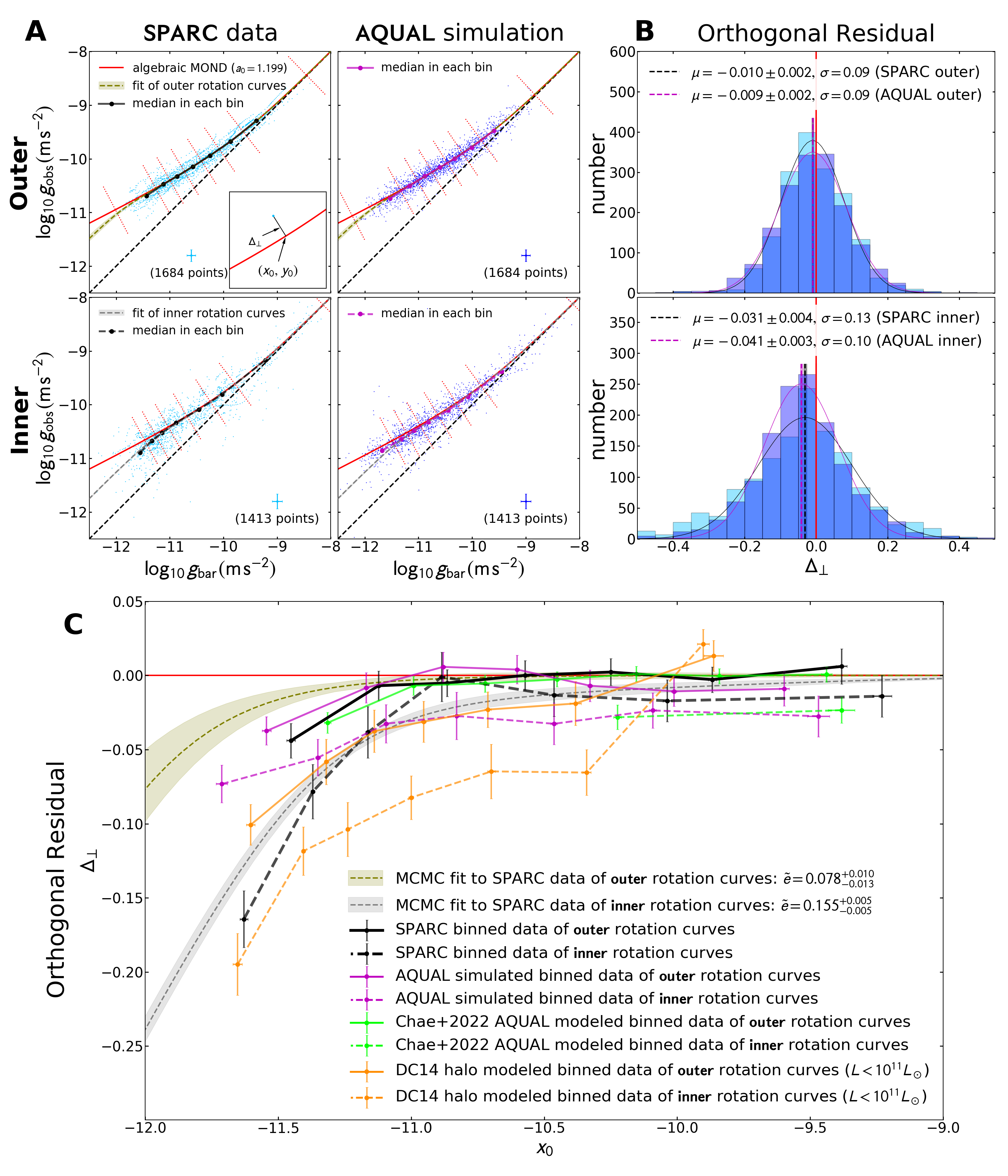}
    \vspace{-0.7truecm}
    \caption{\small 
    The inner and outer parts of rotation curves in the acceleration plane. (A) Each SPARC rotation curve shown in Figure~\ref{RAR_DM}(a) is separated into the inner rising and outer quasi-flat parts. Data for the two parts are displayed separately and compared with the AQUAL simulated results for the same galaxies. Each panel is in the same format as Figure~\ref{RAR_DM}(a). The dashed curve and band indicate the Bayesian fit of Equation~(\ref{eq:aqformula}) to the SPARC data points of the inner or outer part. (B) Histograms show orthogonal residuals from the algebraic MOND relation (red line) for the outer (upper panel) and inner (lower panel) data points. Histograms are fitted with a Gaussian model, and the fitted mean $\mu$ (with a bootstrap error) and standard deviation $\sigma$ are indicated. (C) Solid and dashed lines show the trend of orthogonal residuals in orthogonal bins for the outer and inner parts, respectively. Thick black lines represent the SPARC data while magenta lines represent the AQUAL simulated results. Green lines represent the MCMC fitted results for a subsample of 65 galaxies \citep{chae2022b} of a good dynamic range that are relatively more massive and larger. Orange lines represent the DC14 results for 111 bulgeless galaxies with $L < {10^{11}}{L_\odot }$, which is the most favorable CDM case shown in Figure~\ref{RAR_DM}.
    } 
   \label{RAR_AQUAL}
\end{figure}

Modified gravity and modified inertia can be distinguished by comparing the inner rising part and the outer quasi-flat part of RCs in the acceleration plane. Modified gravity predicts an EFE-unrelated systematic deviation of the inner rising part \citep{brada1995,angus2012,chae2022a} from the algebraic MOND relation in axisymmetric flattened systems while modified inertia does not for circular orbits in axisymmetric systems \citep{milgrom1994}. Figure~\ref{RAR_AQUAL}(a) shows the data separately for the inner and outer parts of the 152 SPARC galaxies and compares them with the predictions of AQUAL. The inner and outer parts are clearly distinguished, in agreement with the prediction of AQUAL. The difference between the inner and outer parts can also be quantified by the parameter $\tilde e$ of Equation~(\ref{eq:aqformula}) fitted to the data points of each part. The Bayesian fit \citep[see Section~\ref{sec:results_corner} below;][]{chae2022b} returns $\tilde e = 0.078_{ - 0.013}^{ + 0.010}$ with ${a_0} = 1.199_{ - 0.022}^{ + 0.023} \times {10^{ - 10}}{\rm{ m }}{{\rm{s}}^{ - 2}}$ (outer) and $\tilde e = 0.155_{ - 0.005}^{ + 0.005}$ with ${a_0} = 1.114_{ - 0.034}^{ + 0.035} \times {10^{ - 10}}{\rm{ m }}{{\rm{s}}^{ - 2}}$ (inner). The inferred values of $a_0$ are consistent, but the inferred $\tilde{e}$ values show a $6.9\sigma $ difference. The orthogonal residual of the inner part of the SPARC data with respect to the algebraic MOND relation ($- 0.031 \pm 0.004$) is also $ \approx 5\sigma $ different from that of the outer part ($ - 0.010 \pm 0.002$) as shown in Figure~\ref{RAR_AQUAL}(b). The AQUAL-predicted residuals are consistent with these SPARC residuals. The higher estimated $\tilde{e}$ in the inner region can be thought of as arising from the vertical gravity just outside the thin disk plane \citep{banik2018}.

Figure~\ref{RAR_AQUAL}(c) further illustrates the trend of the binned medians of orthogonal residuals for each part of the RCs. This clearly shows the difference between the inner and outer parts.  This panel also includes the AQUAL Bayesian fit results for the 65 galaxies whose outer parts were fitted with Equation~(\ref{eq:aqformula}) by \cite{chae2022b}.  For the outer part, the deviation of the SPARC data for ${x_0} \le  - 11$ is consistent with the AQUAL EFE for an external field strength $\tilde e \la 0.1$, where $x_0$ is the $\log_{10}g_\text{bar}$ value of the point projected on the algebraic MOND curve as defined in the figure. For the inner part, the trend of the SPARC data is consistent with the AQUAL prediction and Bayesian fit results. Note that the Bayesian fitted galaxies are relatively more massive and larger so that the inner data cover only ${x_0} \ge  - 10.5$. 

Figure~\ref{RAR_AQUAL}(c) also shows the DC14 modeling results for 111 bulgeless galaxies with $L < {10^{11}}{L_ \odot }$, which is the most favorable CDM case among those shown in Figure~\ref{RAR_DM}. The outer part of the DC14 model deviates from the SPARC outer RAR at a significance of $ \approx 3\sigma $ in terms of fitting the data with Equation~(\ref{eq:aqformula}). Moreover, the inner part deviates by a larger amount from the SPARC inner RAR. 

\subsection{Bayesian Fit of an AQUAL Model to Stacked Data Points of Rotation Curves}  \label{sec:results_corner}

The stacked data $(u\equiv{\log _{10}}{g_{{\rm{bar}}}},v\equiv{\log _{10}}{g_{{\rm{obs}}}})$ shown in Figure~\ref{RAR_AQUAL} are fitted by Equation~(\ref{eq:aqformula}) through a Bayesian MCMC method introduced (and applied to the outer data) by \cite{chae2022b}. The likelihood function for a curve fit $\mathcal{L}_\text{curve-fit}$ is defined by \citep{chae2022b}
\begin{equation}
    \ln\mathcal{L}_\text{curve-fit} =-\frac{1}{2} \sum_i \left( \frac{\Delta_{\bot,i}^2}{s_\theta^2\sigma_{u_i}^2 +c_\theta^2\sigma_{v_i}^2} + \ln[2\pi(s_\theta^2\sigma_{u_i}^2 +c_\theta^2\sigma_{v_i}^2 )] \right),
    \label{eq:Lcurve}
\end{equation}
where $\Delta_{\bot,i}$ is the orthogonal distance of point $(u_i, v_i)$ from the model curve (Equation~(\ref{eq:aqformula}) in the present case) and its error is contributed by $s_\theta\sigma_{u_i}$ and $c_\theta\sigma_{v_i}$ in which $s_\theta\equiv\sin\theta$ and $c_\theta\equiv\cos\theta$ for the angle $\theta$ of the tangent line of the curve from the horizontal axis. Here it is assumed that data points $(u_i,v_i)$ from various RCs are independent and all reported uncertainties are Gaussian. As noted by the SPARC team \citep{mcgaugh2016}, each data point can be considered independent. In the present context it has an independent information for gravity at the radius where $g_{\rm{bar}}$ and  $g_{\rm{obs}}$ are estimated. This assumption is justified under the MOND paradigm because the AQUAL model represented by Equation~(\ref{eq:aqformula}) is a universal model independent of galaxy type and radius.

However, this approach based on Equation~(\ref{eq:Lcurve}) can have a few caveats. First, the SPARC sample of RCs is not homogeneous. Over many years a number of scientists made observations with various telescopes and produced the outputs through various (thus heterogeneous) data reduction pipelines. Consequently, the reported statistical errors are not homogeneous and some RCs may have unquantified systematic errors. The working assumption is that various systematic errors and peculiarities are canceled out by stacking $>100$ galaxies. Second, $g_\text{bar}$ is derived from uniform near-infrared images taken with the Spitzer space telescope and neutral hydrogen (HI) images. However, there may arise some galaxy-specific systematic uncertainties from geometries of stellar and gas disks, stellar mass-to-light ratios, and molecular gases. It is assumed that possible systematic errors are galaxy specific and not in the same direction for all galaxies. Finally, galaxies are under different environments and thus are subject to different external fields. To be precise, a single curve  should be fitted to the stacked data of galaxies that are subject to the same external field strength. It is assumed that the effects of galaxy-specific peculiar external fields are canceled out by stacking $>100$ galaxies.

\begin{figure}
  \centering
  \includegraphics[width=1.\linewidth]{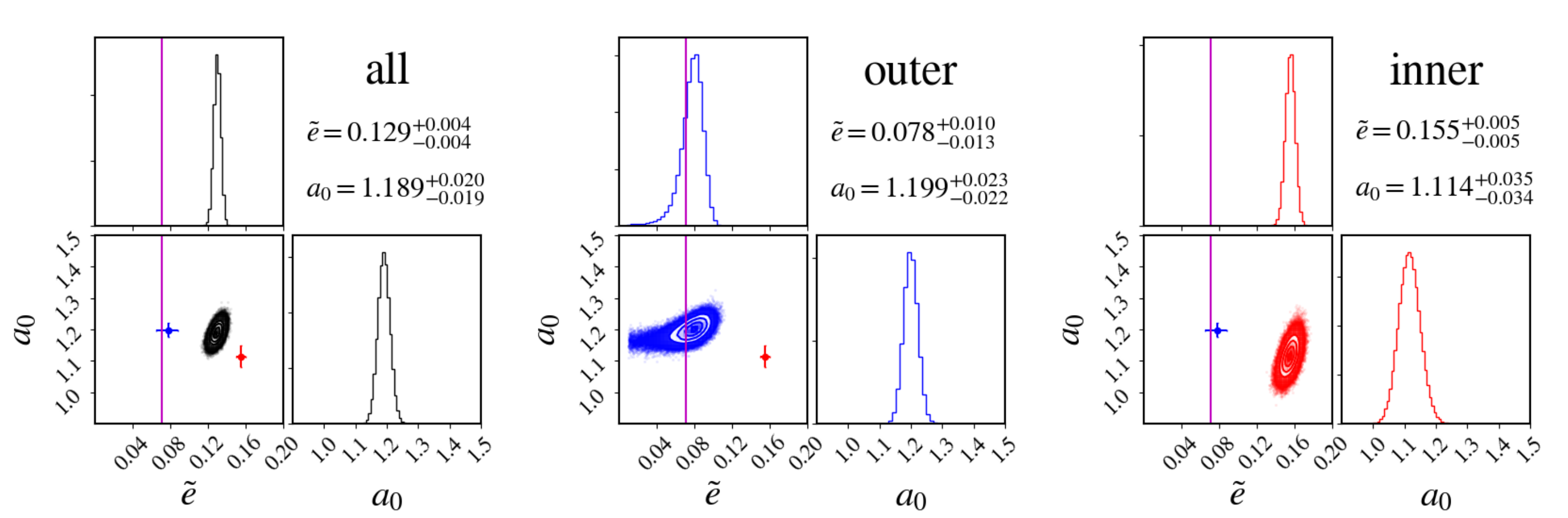}
    \vspace{-0.3truecm}
    \caption{\small 
    Bayesian fit of the AQUAL model (Equation~\ref{eq:aqformula}) to all data points (left), the outer RC data points (center), and the inner RC data points (right). Blue and red points with errorbars indicate the fitted values for the outer and inner parts, respectively. The difference in the fitted values of $\tilde{e}$ for the inner and outer parts is evident, whereas the fitted $a_0$ values are consistent with each other. The magenta vertical line indicates the mean external field expected from the cosmic baryon distribution in the local universe \citep{chae2021,chae2022b}, which is consistent only with the fitted $\tilde{e}$ for the outer part. 
    } 
   \label{corner}
\end{figure} 

Figure~\ref{corner} shows the probability density functions (PDFs) and contour plots of parameters $\tilde e$ and ${a_0}$ for three sets of data points: all points, the points only from the outer part, and the points only from the inner part. These PDFs clearly show that only the value $\tilde e = 0.078_{ - 0.013}^{ + 0.010}$ for the outer points is consistent with the value $\tilde e = 0.071$ (the magenta vertical line) from the large-scale distribution of baryonic matter in the local universe \citep{chae2021}. The unacceptably large value $\tilde e = 0.155 \pm 0.005$ for the inner points cannot be due to the EFE. Moreover, the inferred $\tilde{e}$ for the inner points is somewhat larger if we fix $a_0$ to the value inferred from the outer points. This is because the weaker gravity $g_\text{obs}$ in the inner parts can be partly accounted for in the fit through a lower $a_0$ than for the outer data. These results agree only with the predictions of modified gravity.

\begin{figure}
  \centering
  \includegraphics[width=1.\linewidth]{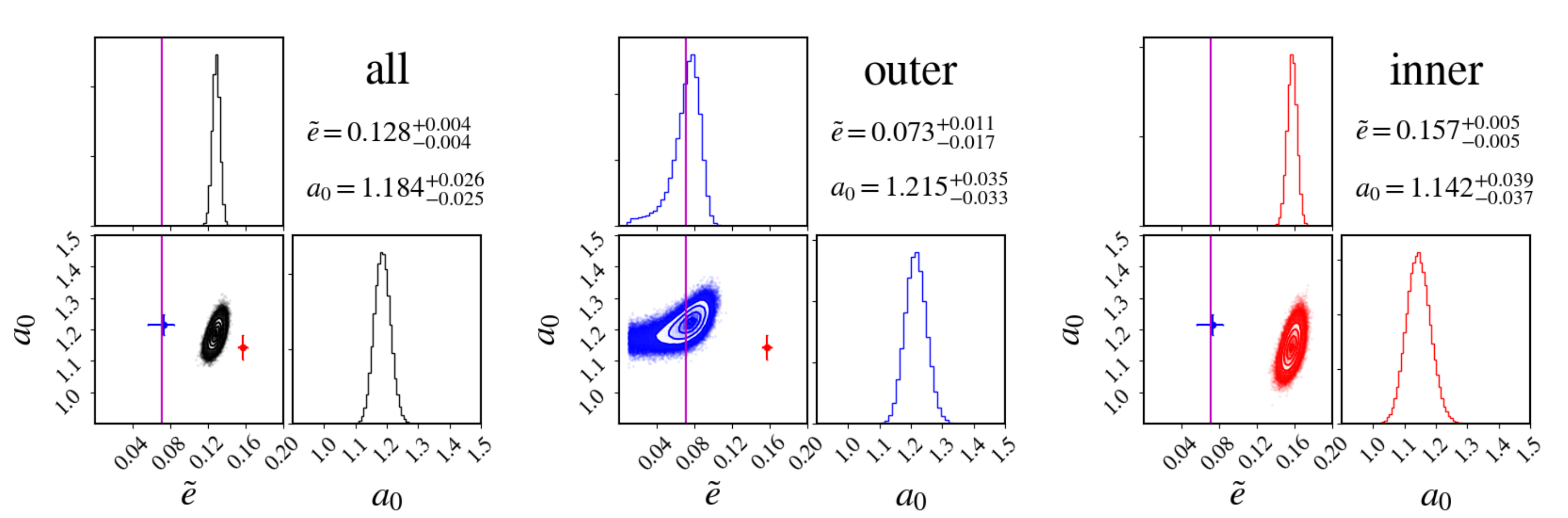}
    \vspace{-0.3truecm}
    \caption{\small 
    Same as Fig.~\ref{corner} but for a subsample of 111 bulgeless galaxies with $L < {10^{11}}{L_\odot }$. The difference between the inner and outer parts remains nearly the same. 
    } 
   \label{corner_noBul}
\end{figure} 

The results shown in Figure~\ref{corner} are for all 152 galaxies selected for this work. Because reported rotation velocities in the inner regions of massive spiral galaxies having a bulge can be less accurate due to the velocity dispersion in the bulge, we also consider a subsample of 111 galaxies without a detected bulge and luminosity $L < {10^{11}}{L_ \odot }$. Figure~\ref{corner_noBul} shows the results. The difference between the inner and outer parts remains nearly the same. Also, the fitted value of $\tilde e = 0.073_{ - 0.017}^{ + 0.011}$ for the outer data points is now in even better agreement with the expected value from the cosmic environment.

\section{Discussion and Conclusions}  \label{sec:discussion}

Modified gravity represented by the AQUAL theory can naturally explain both the inner and outer RARs of the SPARC data with a plausible mean external field strength. Various CDM halo models predicted by recent state-of-the-art hydrodynamic simulations have some difficulty in reproducing the total RAR or the inner-outer separated RARs. In particular, the inner part of CDM results is discrepant with the SPARC inner RAR, even if excluding massive galaxies with a bulge for which BH feedback and measurement errors of rotation velocities are possible concerns. Considering the wide range of halo models tested here, it is unclear whether revised CDM galaxy formation models or alternative DM models such as self-interacting DM \citep{spergel2000} can provide a resolution. However, it is interesting to note that CDM predicts robustly that the inner RAR lies below the outer RAR as in the AQUAL prediction. Also, it is in principle possible that halo models can be fine tuned to match the SPARC inner and outer RARs. What is surprising is that the AQUAL theory can match the RARs naturally without any fine tuning.

The hypothesis that the inner and outer RARs follow a universal curve for axisymmetric systems is ruled out at $5\sigma $ significance if the current SPARC database is taken at face value as demonstrated in Figures~\ref{RAR_AQUAL}, \ref{corner}, \& \ref{corner_noBul}. Even if galaxies with a large luminosity ($L > {10^{11}}{L_ \odot }$) or a bulge are removed from the sample to minimize a possible systematic error, the shift between the inner and outer parts remains nearly the same. One may still consider a more conservative case that galaxies without a detected outer part (i.e.\ galaxies such as those in the bottom row of Figure~\ref{examples})\footnote{These are mostly late-type dwarf galaxies.} are removed further because they do not provide the {\it{control}} outer part within themselves. In this case, we are left with pure-disk galaxies that have both well-defined inner and outer parts. They provide only 607 inner points and 680 outer points. For these relatively small samples, $\tilde{e}$ and $a_0$ cannot be constrained well simultaneously. Thus, if a canonical value of $a_0=1.20$ is assumed and only $\tilde{e}$ is fitted, the two samples give $\tilde{e}=0.140_{-0.008}^{+0.008}$ (inner) and $\tilde{e}=0.069_{-0.014}^{+0.010}$ (outer). These values differ by $\approx 5.5\sigma$. This is a somewhat weaker difference than that for the full sample but still very significant.

To circumvent possible systematic errors one may consider a different galaxy sample. For example, \cite{sanders2019} considered 12 carefully selected gas-dominated galaxies. Because these galaxies are heavily gas dominated, they are free from usual uncertainties in the mass-to-light ratio of stellar disks. The \cite{sanders2019} RCs provide data points of the inner part in the context of this study. When one examines the \cite{sanders2019} RCs along with the overplotted MOND predictions based on the so-called standard IF (rather than the simple IF, or the SPARC RAR, as used here and in the current literature in general), it seems at first sight that the inner points lie, in most cases, well on the algebraic MOND. If this is the case, it would support modified inertia contradicting the results of this study. However, because the standard IF predicts a weaker gravity for $0.1 \la g_\text{bar}/a_0 \la 10$ (see, e.g., figure~1 of \citealt{chae2019}) than the one used in this study, the apparent agreement with the standard IF implies actually that the inner RCs are a little lower than the algebraic MOND represented by the simple IF. Since \cite{sanders2019} did not test quantitatively his selected RCs in the context of distinguishing between modified inertia and modified gravity, it is interesting to consider a similar sample selected using the \cite{sanders2019} criteria from the SPARC database. By requiring that the gas mass is at least 2.2 times larger than the stellar mass and the inclination is in the range $40^\circ \le i < 80^\circ$, we are left with 29 gas-dominated galaxies that provide 336 inner data points. It is found that these data points lie below the algebraic MOND in the average/median sense and the Bayesian fitting gives $\tilde{e}=0.155_{-0.005}^{+0.005}$ (for a fixed $a_0=1.2$ as the data points do not provide high-acceleration data points), which is indistinguishable from the value for the total sample shown in Figure~\ref{corner}. 

Would it be still possible that the SPARC inner and outer parts are made to follow the same RAR supporting modified inertia by some systematic effects? A point on the acceleration plane can be moved vertically or horizontally by adjusting parameters systematically. Thus, it is in principle possible to do a fine tuning of systematic shifts to make inner points match outer points within some tolerable uncertainty. For example, consider the above conservative sample of pure disks with 607 inner points and 680 outer points (this sample is easier to do fine tuning than the full sample). A systematic decrease of 15\% in the $V_\text{bar}$ of inner points accompanied by a smaller systematic decrease in the $V_\text{obs}$ would bring them to about $1.5\sigma$ agreement with outer points. However, although such systematic shifts may not be impossible in principle, the author is not aware of any plausible scenario for such a one-sided systematic shift (rather than galaxy-specific shifts).

A recent analysis of 15 dwarf galaxies to distinguish between modified inertia and modified gravity \citep{petersen2020} based on a dimensionless global parameter \citep{milgrom2012} obtained a marginal preference for modified gravity. This work clearly shows that modified gravity is preferred over modified inertia. Ironically, if the current SPARC database suffered from a gross systematic error so that the inner and outer RARs were universal, both CDM and modified gravity would be ruled out and modified inertia would be favored.  However, as shown above, the current SPARC database prefers modified gravity represented by the AQUAL theory over modified inertia or CDM. This may be the desired result for MOND because modified gravity can be well understood at least nonrelativistically through Lagrangian theories while it seems a tall order to build a modified inertia theory (see section~2.5 of \citealt{banik2022} for a discussion).

When the SPARC database was analyzed by \cite{chae2020b,chae2021} with an analytic spherical MOND model from \cite{famaey2012}, it was a surprise that external fields inferred from internal dynamics based on such a simple model agreed with those expected from cosmic environments. Those previous studies called for further studies based on more accurate and realistic modeling \citep{chae2022a,chae2022b}. Now that the inner and outer parts of RCs are separated and numerical solutions of modified gravity are used here, the EFE interpretation of SPARC RCs is upheld. In this regard, it is interesting to note a recent study supporting EFE in dwarf (spheroidal and elliptical) galaxies of the Fornax Cluster \citep{asencio2022}. It shows that the EFE from the cluster needs to be considered when finding the tidal radii of the dwarf galaxies, as otherwise they would have a lot more self-gravity, making them immune to tides, thus contradicting observations. 

In conclusion, from $\Lambda$CDM and MOND Bayesian analyses and AQUAL numerical simulations of the inner and outer parts of 152 good-quality SPARC RCs, the following are found: 
\begin{itemize}
    \item Under a plausible cosmic mean external field, the AQUAL modified gravity model predicts correctly both the inner and outer parts of RCs.
    \item When various Bayesian modeling results are tested by the orthogonal residual $\Delta_\bot$ on the acceleration plane, the current $\Lambda$CDM halo models tend to deviate outside the range allowed by the systematic uncertainty of stellar mass-to-light ratios.
    \item When the $\Lambda$CDM halo models are compared with the AQUAL model through the Bayesian information criterion $\Delta$BIC, there is a positive evidence for the AQUAL model.
    \item Taken at face value, the current SPARC database rules out at $7\sigma$ significance the hypothesis that the inner and outer parts follow a universal curve on the acceleration plane. This would contradict current proposals of modified inertia.
    \item The EFE interpretation of galactic rotation curves is upheld and its origin is likely to be modified gravity rather than modified inertia.
\end{itemize}

\section*{Acknowledgments}
The author thanks S. McGaugh, F. Lelli, and J. Schombert for their help with understanding the SPARC database, and M. Milgrom for insightful communications. It is the author's pleasure to acknowledge a very useful anonymous report that helped improve the presentations significantly. The manuscript also benefited from a review by a Statistics Editor. This work was supported by the National Research Foundation of Korea (grant No. NRF-2022R1A2C1092306). This work was also in part supported by the faculty research fund of Sejong University in 2022.

\bibliographystyle{aasjournal}

\newpage

\appendix

\section{Methods} \label{sec:appA}

\subsection{Bayesian Modeling}

A Bayesian approach is used to fit a model to the observed the rotation curve of a galaxy. The method is essentially the same as that adopted in \cite{chae2020b,chae2022b} in their MOND modeling. Here both DM halo and MOND models are considered. A likelihood $\mathcal{L}$ for a given model with free parameters ${\bm{\beta }} = \{ {\beta _k}\} $, which includes inclination $i$, is defined as\footnote{The second logarithmic term in the bracket was actually used by \cite{chae2020b} who however did not state it explicitly.}	
\begin{equation}
    \ln\mathcal{L} =  - \frac{1}{2}\sum\limits_{j = 1}^N {\left[ {{{\left( {\frac{{{V_{{\rm{rot}}}}(i;{R_j}) - {V_{{\rm{mod}}}}({\bm{\beta }};{R_j})}}{{{\sigma _{{V_{{\rm{rot}}}}(i;{R_j})}}}}} \right)}^2} + \ln \left( {2\pi \sigma _{{V_{{\rm{rot}}}}(i;{R_j})}^2} \right)} \right]},
    \label{eq:S1}
\end{equation}	
where ${V_{{\rm{rot}}}}(i;{R_j}) = {V_{{\rm{obs}}}}({R_j})\sin ({i_{{\rm{obs}}}})/\sin (i)$, ${\sigma _{{V_{{\rm{rot}}}}(i;{R_j})}} = {\sigma _{{V_{{\rm{obs}}}}({R_j})}}\sin ({i_{{\rm{obs}}}})/\sin (i)$, and ${V_{{\rm{mod}}}}({\bm{\beta }};{R_j})$ is the theoretical velocity that depends on the model. Then, the posterior probability of ${\bm{\beta }}$ is given by
\begin{equation}
    p({\bm{\beta }}) = {\cal L} \times \prod\limits_k {{\rm{Pr}}} ({\beta _k}),
    \label{eq:S2}
\end{equation}
where ${\rm{Pr}}({\beta _k})$ is the prior probability of parameter ${\beta _k}$. (See below for prior constraints on inclination and other parameters.) With this definition of probability, the public MCMC package emcee \citep{emcee} is used to derive the posterior PDF of the model parameters.

The theoretical rotation velocity ${V_{{\rm{mod}}}}({\bm{\beta }};{R_j})$ is defined under the DM or MOND hypothesis. Under the DM hypothesis, the theoretical velocity is defined as
\begin{equation}
    {V_{{\rm{mod}}}} = \sqrt {V_{{\rm{mod,bar}}}^2 + V_{{\rm{mod,halo}}}^2},
    \label{eq:S3}
\end{equation}
where ${V_{{\rm{mod,halo}}}}$ is the Newtonian velocity contributed by the DM halo model, and ${V_{{\rm{mod,bar}}}}$ is the Newtonian velocity contributed by the observed baryons, which depends on some of the free parameters as
\begin{equation}
    {V_{{\rm{mod,bar}}}} = \sqrt {\hat D({{\hat \Upsilon }_{{\rm{disk}}}}V_{{\rm{disk}}}^2 + {{\hat \Upsilon }_{{\rm{bulge}}}}V_{{\rm{bulge}}}^2 + {{\hat \Upsilon }_{{\rm{gas}}}}{V_{{\rm{gas}}}}|{V_{{\rm{gas}}}}|)},
    \label{eq:S4}
\end{equation}
where $\hat D \equiv D/{D_{{\rm{obs}}}}$ is the parameter representing the distance to the galaxy normalized by the SPARC-reported distance ${D_{{\rm{obs}}}}$, and ${\hat \Upsilon _{{\rm{disk}}}}$, ${\hat \Upsilon _{{\rm{bulge}}}}$ and ${\hat \Upsilon _{{\rm{gas}}}}$ are the parameters representing galaxy-specific mass-to-light ratios and the gas-to-HI mass ratio normalized by 0.5,  0.7, and 1.33, respectively. Note that ${V_{{\rm{gas}}}}$ can make a negative contribution in some rare cases in the inner region of the galaxy due to deviations from spherical symmetry \citep{lelli2016}.

Under the MOND hypothesis, the theoretical velocity is given by
\begin{equation}
    {V_{{\rm{mod}}}} = \sqrt {\frac{{{g_{{\rm{MOND}}}}}}{{{g_{{\rm{bar}}}}}}} {V_{{\rm{bar}}}},
    \label{eq:S5}
\end{equation}
where ${g_{{\rm{MOND}}}}/{g_{{\rm{bar}}}}$ specifies the MOND model and is given by Equation~(\ref{eq:aqformula}) in the case of AQUAL.

\subsection{The Approach of this Work to CDM Halos}

A number of studies in the literature compared statistical properties of $\Lambda {\rm{CDM}}$-simulated mock galaxies (e.g.\ \citealt{dicintio2016,desmond2017,keller2017,ludlow2017,navarro2017,tenneti2018,paranjape2021a,paranjape2021b}) with the SPARC RAR. Here we build halo models for actual SPARC galaxies based on recent hydrodynamic simulations of galaxy formation in $\Lambda {\rm{CDM}}$. This approach is similar to that used in \cite{li2020} where a number of halo models were built for SPARC galaxies. However, here only sophisticated halo models with constraints from recent hydrodynamic simulations are considered.

Current state-of-the-art hydrodynamic simulations of galaxy formation under the $\Lambda {\rm{CDM}}$ cosmological model predict density profiles of CDM halos. Here we need simulation results for relatively low-mass halos with ${M_{{\rm{halo}}}} \mathbin{\lower.3ex\hbox{$\buildrel<\over
{\smash{\scriptstyle\sim}\vphantom{_x}}$}} {10^{13}}{M_\odot }$ applicable to most SPARC galaxies. We consider results from three independent projects: the NIHAO series \citep{tollet2016,dekel2017,maccio2020}, the DC14 analyses of their simulations \citep{dicintio2014a,dicintio2014b}, and the FIRE-2 simulation \citep{lazar2020}. The details of these results are described below. The DC14 results were also used in a previous MCMC analysis of CDM halos \citep{katz2017}. That study stressed that the DC14 model matched observed galaxies better than the NFW model based on DM-only simulations. However, they did not test the DC14 model based on a statistical RAR (let alone the inner-outer separation of RCs) as done here.

\subsection{CDM Halo Masses and Radii}

The radius of a halo denoted by ${r_{200}}$ is defined as the radius within which the average mass density is 200 times the present-day critical density of the universe ${\rho _{{\rm{crit,0}}}} = 3H_0^2/(8\pi G)$, where ${H_0}$ is the Hubble constant. The mass of the halo denoted by ${M_{200}}$ is the mass bounded by ${r_{200}}$. Then, the numerical value of ${r_{200}}$ is given by
\begin{equation}
    {r_{200}} = 0.02063{\left(\frac{M_{200}}{{\rm{M}_\odot }}\right)^{1/3}}h_{70}^{ - 2/3}{\rm{ kpc}},
    \label{eq:S6}
\end{equation}
where ${h_{70}} \equiv {H_0}/(70\hspace{1ex}{\rm{ km}}\hspace{1ex}{{\rm{s}}^{ - 1}} \hspace{1ex}{\rm{ Mpc}^{ - 1}})$. Throughout this work the value of ${H_0}$ is taken to be $73\hspace{1ex}{\rm{ km }}\hspace{1ex}{{\rm{s}}^{ - 1}}\hspace{1ex}{\rm{ Mpc}^{ - 1}}$ and so ${h_{70}} = 73/70$. 

Because some authors use different definitions of halo radius and mass than the above, we also consider virial radius and mass (${r_{{\rm{vir}}}}$ and ${M_{{\rm{vir}}}}$) from \cite{bryan1998}. In this definition, the overdensity factor for the halo region is about one half of 200, so that ${r_{{\rm{vir}}}} \approx 1.4{r_{200}}$ and ${M_{{\rm{vir}}}}$ is accordingly related to ${M_{200}}$.

We use the popular abundance matching relation between halo mass ${M_{200}}$ and stellar mass ${M_ \star }$ to provide a constraint on the halo. The following relation is taken from the literature
\begin{equation}
    \log_{10}M_\star = \log_{10}(\epsilon M_1) + f\left[\log_{10}\left(\frac{M_{200}}{M_1}\right)\right] - f(0),
    \label{eq:S7}
\end{equation}
where the function $f(x)$ is defined as
\begin{equation}
    f(x) =  - {\log _{10}}({10^{ - ax}} + 1) + \frac{\delta{{{[{{\log }_{10}}(1 + {\mathrm{e}^x})]}^b}}}{{1 + {\mathrm{e}^{{{10}^{ - x}}}}}}.
    \label{eq:S8}
\end{equation}
We consider two results \citep{behroozi2013,kravtsov2018} that bracket the current range. \cite{behroozi2013} report $a = 1.412$, $b = 0.316$, $\delta  = 3.508$, ${\log _{10}}{M_1/{\rm{M}}_\odot} = 11.514$, and $\log_{10}\epsilon=-1.777$ while \cite{kravtsov2018} report $a = 1.779$, $b = 0.547$, $\delta  = 4.394$, ${\log _{10}}{M_1/{\rm{M}}_\odot} = 11.35$, and $\log_{10}\epsilon=-1.642$. These two differ appreciably only for large masses ${M_{200}} \mathbin{\lower.3ex\hbox{$\buildrel>\over
{\smash{\scriptstyle\sim}\vphantom{_x}}$}} {10^{12}}{M_\odot }$. We will see below that the posterior ${M_{200}}$-${M_ \star }$ relation is closer to the \cite{kravtsov2018} relation, consistent with an earlier finding \citep{dicintio2016}. The parameter ${\rho _{\rm{s}}}$ in Equation~(\ref{eq:zhao}) is determined by ${M_{200}}$. 

\subsection{Rotation Velocity Contributed by the Halo}

For a halo described by Equation~(\ref{eq:zhao}), the halo contribution to the rotation velocity ${V_{{\rm{mod,halo}}}}$ is given by	
\begin{equation}
    {V_{{\rm{mod,halo}}}}(R) = {10^3}{\left[ {\frac{{4.3{M_{200}}/({{10}^{12}}{{\rm{M}}_ \odot })}}{{R/{\rm{kpc}}}} \times \frac{{f({c_{200}}R/{r_{200}},\alpha ,\beta ,\gamma )}}{{f({c_{200}},\alpha ,\beta ,\gamma )}}} \right]^{1/2}}{\rm{km }}\hspace{1ex}{{\rm{s}}^{ - 1}},
    \label{eq:S9}
\end{equation}
with the definition 
\begin{equation}
    f(x,\alpha ,\beta ,\gamma ) \equiv \int_0^x {\frac{{{s^{2 - \gamma }}}}{{{{(1 + {s^\alpha })}^{(\beta  - \gamma )/\alpha }}}}} ds.
    \label{eq:S10}
\end{equation}
In Equation~(\ref{eq:S9}), we use the usual concentration parameter ${c_{200}} \equiv {r_{200}}/{r_{\rm{s}}}$ although its meaning is different than that of DM-only simulations. For the core-Einasto profile \citep{lazar2020}, the integral function of Equation~(\ref{eq:S10}) is modified accordingly.

 \begin{table}
\caption{Summary of the Model Parameters and Prior Constraints}\label{tab:prmt}
\begin{center}
  \begin{tabular}{lcccc}
  \hline
 Parameter & Relevant Model & Distribution & $(\mu,\sigma)$ or Range or Value & Comment \\
 \hline
$\log_{10}(\hat{\Upsilon}_\text{disk})$ & All & Gaussian  & $(0, 0.1)$ &  $\hat{\Upsilon}_\text{disk}\equiv \Upsilon_\text{disk}/0.5$ \\
$\log_{10}(\hat{\Upsilon}_\text{bulge})$ & All & Gaussian  & $(0, 0.1)$ &  $\hat{\Upsilon}_\text{bulge}\equiv \Upsilon_\text{bulge}/0.7$ \\
$\log_{10}(\hat{\Upsilon}_\text{gas})$ & All & Gaussian  & $(\log_{10}(X^{-1}/1.33), 0.1)^\dagger$ &  $\hat{\Upsilon}_\text{gas}\equiv \Upsilon_\text{gas}/1.33$  \\
$\log_{10}(\hat{D})$ & All & Gaussian  & $(0, \log_{10}(1+\sigma_{D_\text{obs}}/D_\text{obs}))$ &  \\
$i$      & All  &  Gaussian   &  $(i_\text{obs},\sigma_{i_\text{obs}})$ &  \\
$\log_{10}(M_{200}/{\rm M}_\odot)$ & DC14/NIHAO/FIRE-2  & N/A  &  N/A  &  By Equation~(\ref{eq:S7}) \\
\multirow{2}{*}{$c_{200}$}  &  NIHAO  &  N/A  &  N/A  &  Unconstrained  \\
                           & DC14  &  N/A  &  N/A  &  By Equations~(\ref{eq:S13}) \& (\ref{eq:S14})  \\
    $\tilde{c}_{200}$      & FIRE-2  &  N/A  &  N/A  &  Unconstrained  \\
\multirow{2}{*}{$\alpha$}  &  NIHAO  &  Fixed  &  0.5  &   \\
                           & DC14  &  Gaussian  &  (Equation~(\ref{eq:S12}),0.05)  &   \\
\multirow{2}{*}{$\beta$}  &  NIHAO  &   Fixed  &  3.5  &   \\
                          &  DC14   &  Gaussian  &  (Equation~(\ref{eq:S12}),0.05)  &   \\
\multirow{2}{*}{$\gamma$}  &  NIHAO  &  N/A      &  N/A          & By Equation~(\ref{eq:S11}) \\
                          &  DC14   &  Gaussian  &  (Equation~(\ref{eq:S12}),0.05) &  \\
  $\alpha_\epsilon$  &  FIRE-2   &  Gaussian  & $(0.16,0.01)$ &  \\
  $r_c$  &  FIRE-2   &  Fixed  & Equation~(\ref{eq:S16}) &  \\
  $\log_{10}(\tilde{e})$  &  AQUAL &  Uniform  & $[-2.6,-0.15]$ & From \cite{chae2022b}   \\
  $a_0$                   &  AQUAL &  Gaussian & $(1.24,0.14)$  & From \cite{chae2022b}   \\
 \hline
\end{tabular}
\end{center}
\hspace{2em} $^\dagger X$ from \cite{mcgaugh2020}  \\
\end{table}

\subsection{Hydrodynamic Simulation Results on Halo Density Profiles: NIHAO}

An analysis of NIHAO halos \citep{dekel2017} found that the simple choice $(\alpha ,\beta ) = (0.5,3.5)$ in Equation~(\ref{eq:zhao}) can be as good as allowing them to be free in describing the simulated halos. Another analysis \citep{lazar2020} from the NIHAO project measured the logarithmic slope between $r = 0.01{r_{200}}$ and $0.02{r_{200}}$. The reported logarithmic slope (multiplied by $ - 1$) over this radial range, denoted by ${\gamma _1}$, is given as a function of $x \equiv {M_ \star }/{M_{200}}$ by
\begin{equation}
    {\gamma _1}(x) = 0.158 + {\log _{10}}\left[ {26.49{{\left( {1 + \frac{x}{{9.44 \times {{10}^{ - 5}}}}} \right)}^{ - 0.85}} + {{\left( {\frac{x}{{8.77 \times {{10}^{ - 3}}}}} \right)}^{1.66}}} \right]
    \label{eq:S11}
\end{equation}
with a $1\sigma $ scatter of 0.18. Equation~(\ref{eq:S11}) is used as a prior constraint on the halo profile.

In this model, the free parameters in Equation~(\ref{eq:S1}) are ${\bm{\beta }} = \{ i,\hat D,{\hat \Upsilon _{{\rm{disk}}}},({\hat \Upsilon _{{\rm{bulge}}}},){\hat \Upsilon _{{\rm{gas}}}},{M_{200}},{c_{200}},\gamma \} $. The priors on $i,\hat D,{\hat \Upsilon _{{\rm{disk}}}},{\hat \Upsilon _{{\rm{bulge}}}}$, and ${\hat \Upsilon _{{\rm{gas}}}}$ are the same as table 1 of \cite{chae2020b}. Additional priors on the DM halo are provided by Equation~(\ref{eq:S7}) and Equation~(\ref{eq:S11}). In the case of Equation~(\ref{eq:S7}), a lognormal prior on ${M_\star }$\footnote{The stellar mass is obtained by ${M_ \star } = {\hat \Upsilon _{{\rm{disk}}}}0.5({L_{{\rm{tot}}}} - {L_{{\rm{bulge}}}}) + {\hat \Upsilon _{{\rm{bulge}}}}0.7{L_{{\rm{bulge}}}}$, where ${L_{{\rm{tot}}}}$ is the total luminosity and ${L_{{\rm{bulge}}}}$ is the luminosity of the bulge. The values of ${L_{{\rm{bulge}}}}$ are taken from the SPARC website: http://astroweb.cwru.edu/SPARC/.} is imposed with a scatter of 0.25 dex. 

The parameters of the NIHAO-based galaxy model and the prior constraints are summarized in Table~\ref{tab:prmt}. The table also includes the parameters and prior constraints of other models including the AQUAL-based model.

\subsection{Hydrodynamic Simulation Results on Halo Density Profiles: DC14}

A suite of cosmological hydrodynamic simulations called Making Galaxies In a Cosmological Context  \citep[MaGICC;][]{brook2012,stinson2013} provides a detailed prediction of the halo density profiles \citep{dicintio2014b} affected by the baryonic physics of galaxy formation. The results presented by \citet[][DC14 model]{dicintio2014b} predict the behaviors of all three shape indices $\alpha $, $\beta $, and $\gamma $ as functions of ${M_ \star }/{M_{{\rm{vir}}}}$. They are
\begin{equation}
    \begin{array}{*{20}{l}}
\alpha & = &{2.94 - {{\log }_{10}}[{{({{10}^{X + 2.33}})}^{ - 1.08}} + {{({{10}^{X + 2.33}})}^{2.29}}]},\\
\beta & = &{4.23 + 1.34X + 0.26{X^2}},\\
\gamma & = &{ - 0.06 + {{\log }_{10}}[{{({{10}^{X + 2.56}})}^{ - 0.68}} + {{10}^{X + 2.56}}]},
\end{array}
    \label{eq:S12}
\end{equation}
where $X \equiv {\log _{10}}({M_ \star }/{M_{{\rm{vir}}}})$ and their definition of ${M_{{\rm{vir}}}}$ is the mass of a sphere within which the mass density is 93.6 times the critical density (typically $1.1{M_{200}} \mathbin{\lower.3ex\hbox{$\buildrel<\over
{\smash{\scriptstyle\sim}\vphantom{_x}}$}} {M_{{\rm{vir}}}} \mathbin{\lower.3ex\hbox{$\buildrel<\over
{\smash{\scriptstyle\sim}\vphantom{_x}}$}} 1.4{M_{200}}$). These scaling relations are used as priors with a small scatter of 0.05. We also require $\alpha \ge 0.5$ because the simulation results do not actually include any value less than 0.5.

Furthermore, the DC14 model predicts how the scale radius ${r_{\rm{s}}}$ depends on a galaxy's star formation efficiency. The radius where the logarithmic slope of the density is $ - 2$, to be denoted by ${r_{ - 2}}$, has the following scaling relation with the above-defined $X$ for the simulated halos
\begin{equation}
    \frac{c_{ - 2}}{c_{{\rm{NFW}}}} = 1.0 + 0.00003{\mathrm{e}^{3.4(X + 4.5)}},
    \label{eq:S13}
\end{equation}
where ${c_{ - 2}} \equiv {r_{{\rm{vir}}}}/{r_{ - 2}} = ({r_{{\rm{vir}}}}/{r_{\rm{s}}}){[(2 - \gamma )/(\beta  - 2)]^{ - 1/\alpha }}$ and ${c_{{\rm{NFW}}}}$ is the concentration of the NFW halos from DM-only simulations. For the NFW concentration we consider \cite{dutton2014} based on the Planck cosmology which works slightly better than \cite{maccio2008} based on the WMAP5 cosmology in matching the SPARC data. The Planck NFW concentration is given by 
\begin{equation}
    {\log _{10}}{c_{{\rm{NFW}}}} = 0.905 - 0.101{\rm{ }}{\log _{10}}\left(\frac{M_{200}}{1.43 \times {10^{12}}h_{70}^{ - 1}{{\rm{M}}_\odot }}\right).
    \label{eq:S14}
\end{equation}
We use Equation~(\ref{eq:S13}) along with Equation~(\ref{eq:S14}) as a prior constraint with a lognormal scatter of 0.1. Thus, the DC14 model has free parameters ${\bm{\beta }} = \{ i,\hat D,{\hat \Upsilon _{{\rm{disk}}}},({\hat \Upsilon _{{\rm{bulge}}}},){\hat \Upsilon _{{\rm{gas}}}},{M_{200}},{c_{200}},\alpha ,\beta ,\gamma \}$, with additional constraints provided by Equations~(\ref{eq:S12}), (\ref{eq:S13}) and (\ref{eq:S14}). When these constraints are used, a proper transformation between ${M_{{\rm{vir}}}}$ (${c_{{\rm{vir}}}}$) and ${M_{200}}$ (${c_{200}}$) is used.
 
\subsection{Hydrodynamic Simulation Results on Halo Density Profiles: FIRE-2}

The Feedback In Realistic Environments (FIRE)-2 simulation results \citep{lazar2020} have been fitted with the “core-Einasto” profile given as
\begin{equation}
    \rho_\text{cEin}(r)=\tilde{\rho}_s \exp\left\{-\frac{2}{\alpha_\epsilon}\left[\left(\frac{r+r_c}{\tilde{r}_s}\right)^{\alpha_\epsilon}-1\right]\right\},
    \label{eq:S15}
\end{equation}
where $r_c$ and $\tilde{r}_s$ are the core radius and scale radius defined for this profile. They find $\alpha_\epsilon=0.16$ and a core radius as a function of the stellar-to-halo mass ratio given as	
\begin{equation}
    {r_c}(x) = {10^{{A_1}}}{\left( {{A_2} + \frac{x}{{x_1^ * }}} \right)^{ - {\beta _1}}}{\left( {\frac{x}{{x_2^ * }}} \right)^{{\gamma _1}}}{\rm{ kpc}}
    \label{eq:S16}
\end{equation}
where $x \equiv {M_ \star }/{M_{{\rm{vir}}}}$, ${A_1} = 1.21$, ${A_2} = 0.71$, $x_1^ *  = 7.2 \times {10^{ - 3}}$, $x_2^ *  = 0.011$, ${\beta _1} = 2.31$, and ${\gamma _1} = 1.55$. The FIRE-2 model has free parameters ${\bm{\beta }} = \{ i,\hat D,{\hat \Upsilon _{{\rm{disk}}}},({\hat \Upsilon _{{\rm{bulge}}}},){\hat \Upsilon _{{\rm{gas}}}},{M_{200}},{\tilde c_{200}}\}$, where ${\tilde c_{200}} \equiv {r_{200}}/{\tilde r_s}$. When Equation~(\ref{eq:S16}) is used, a proper transformation between ${M_{{\rm{vir}}}}$ and ${M_{200}}$ is applied.

\subsection{AQUAL Simulation}

There are two nonrelativistic Lagrangian theories of MONDian modified gravity that can be used to model the galactic rotation curve of a rotationally supported galaxy. They are the AQUAL \citep{bekenstein1984} and QUMOND \citep{milgrom2010} field equations. These modified Poisson’s equations are nonlinear and must be solved numerically. Solutions of the field equations are unusually difficult when an external field is present in an arbitrary direction. Recently, \cite{chae2022a} have investigated numerical solutions of both the AQUAL and QUMOND field equations when an external field is either aligned or inclined with the rotation axis of disks with different thicknesses. QUMOND numerical solutions were also investigated by other researchers \citep{oria2021,zonoozi2021}.

\cite{chae2022b} tested AQUAL and QUMOND theories by comparing an external field fitted to the stacked data of the outer part of rotation curves of 114 SPARC galaxies\footnote{Some rotation curves do not have a noticeable outer part. Thus, \cite{chae2022b} used only 114 galaxies.} with a cosmic mean external field estimated directly from observations of the local universe \citep{chae2021}. \cite{chae2022b} find that AQUAL is preferred over QUMOND by the outer part of rotation curves. The AQUAL-fitted value of the mean external field strength is $\tilde e = 0.078_{ - 0.013}^{ + 0.010}$, which is consistent with the value $\tilde e = 0.071 \pm 0.001$ \citep{chae2021} from the environment for the case that all intergalactic baryons are maximally correlated with visible galaxies (their ``max clustering'' case). Thus, in this work simulations are carried out under the AQUAL theory with $\tilde e = 0.078$.  

A galaxy is represented by a combination of a stellar disk, a gas disk, and a bulge if present. The base model for a disk is the Miyamoto-Nagai \citep[MN;][]{Miyamoto1975} model whose normalized mass distribution is given by
\begin{equation}
    {\rho _{{\rm{MN}}}}(R,z) = \frac{{{b^2}}}{{4\pi }}\frac{{a{R^2} + (a + 3\sqrt {{z^2} + {b^2}} ){{(a + \sqrt {{z^2} + {b^2}} )}^2}}}{{{{[{R^2} + {{(a + \sqrt {{z^2} + {b^2}} )}^2}]}^{5/2}}{{({z^2} + {b^2})}^{3/2}}}},
    \label{eq:S17}
\end{equation}
where $a$ and $b$ are the MN scale parameters in the radial and vertical directions. The stellar disk is represented by an approximate exponential disk with a finite thickness composed of three MN models \citep{smith2015}. The effective mass density of this disk composed of three MN components may be approximated by a double exponential model	
\begin{equation}
    {\rho _{{\rm{disk}}}}(R,z) = {\rho _0}\exp ( - R/{R_{\rm{d}}})\exp ( - |z|/{h_z}),
    \label{eq:S18}
\end{equation}
where ${R_{\rm{d}}}$ and ${h_z}$ are the scale radius and height. Three MN models have different values of $a$ but share a common value of $b$.  For $b/{R_{\rm{d}}} \ll 1$, the scale height is related to $b$ by ${h_z} \approx 0.8b$. For the model disk we choose $b/{R_{\rm{d}}} = 0.1$ so that ${h_z}/{R_{\rm{d}}} \approx 0.08$ consistent with observations of typical disk galaxies\footnote{Observed rotationally supported galaxies can be thinner or thicker than ${h_z}/{R_{\rm{d}}} = 0.08$. However, the effect of varying ${h_z}/{R_{\rm{d}}}$ is minor unless the galaxy becomes close to the spherical shape \citep{chae2022a}.}. The gas disk is simply represented by a single MN model with $a = 2{R_{\rm{d}}}$ and $b = 0.1a$. 

The bulge is represented by the Hernquist model \citep{hernquist1990}:
\begin{equation}
    {\rho _{\rm{H}}}(r) = \frac{{{M_{{\rm{bulge}}}}{r_{\rm{H}}}}}{{2\pi r{{(r + {r_{\rm{H}}})}^3}}},
    \label{eq:S19}
\end{equation}
where ${r_{\rm{H}}}$ is the Hernquist scale radius. Following \cite{li2022} we use the empirical relation
\begin{equation}
    {\log _{10}}({r_{\rm{H}}}) = 1.29{\log _{10}}({R_{\rm{d}}}) - 0.77.
    \label{eq:S20}
\end{equation}

The mass of the stellar disk (${M_{ \star {\rm{disk}}}}$) is given by ${M_{ \star {\rm{disk}}}}/{M_ \odot } = 0.5({L_{{\rm{tot}}}} - {L_{{\rm{bulge}}}})/{L_\odot }$, where ${L_{{\rm{tot}}}}$ and ${L_{{\rm{bulge}}}}$ are the total and bulge luminosities, respectively \citep{lelli2016}. The stellar mass of the bulge is given by ${M_{ \star {\rm{bulge}}}}/{M_\odot } = 0.7{L_{{\rm{bulge}}}}/{L_\odot }$.  The mass of the gas disk is given by ${M_{{\rm{gas}}}} = 1.36{M_{{\rm{HI}}}}$ where ${M_{{\rm{HI}}}}$ is the mass of neutral hydrogen \citep{lelli2016} and the factor 1.36 accounts for other gases, especially primordial helium \citep{chae2020b,mcgaugh2020}.

According to the study by \cite{chae2022a}, the azimuthally averaged radial acceleration depends on the strength of the external field but has little dependence on its inclination. On the other hand, the radial acceleration in the inner part is essentially fixed by the disk geometry. A typical external field ($\tilde e \mathbin{\lower.3ex\hbox{$\buildrel<\over
{\smash{\scriptstyle\sim}\vphantom{_x}}$}} 0.1$) has a minor secondary effect both in AQUAL and QUMOND. It is thus sufficient to consider the axisymmetric case that the external field is aligned with the rotation axis in predicting the azimuthally averaged radial acceleration in rotation curves. We use the AQUAL numerical code \citep{chae2022a} developed recently based on the algorithm proposed by \cite{milgrom1986}.  Here the code is adjusted to include the multiple components of a galaxy, all of which are analytical. The code returns simulated $({g_{{\rm{bar}}}},{g_{{\rm{obs}}}})$ at the observed radii in each rotation curve of 152 SPARC galaxies. Note that azimuthal averaging is not required as the model is axisymmetric in the \cite{chae2022a} code.

\section{Posterior Properties of Dark Matter Halos}  \label{sec:appB}

Bayesian MCMC modeling has returned posterior PDFs of all model parameters. Here posterior values of the halo parameters are compared with their prior values. Also, posterior values of the DM halo density slope between 1\% and 2\% of ${r_{200}}$ are compared with predictions from various cosmological hydrodynamic simulations. 

\subsection{${M_\star} - {M_{200}}$ Relation}

\begin{figure}
  \centering
  \includegraphics[width=1.\linewidth]{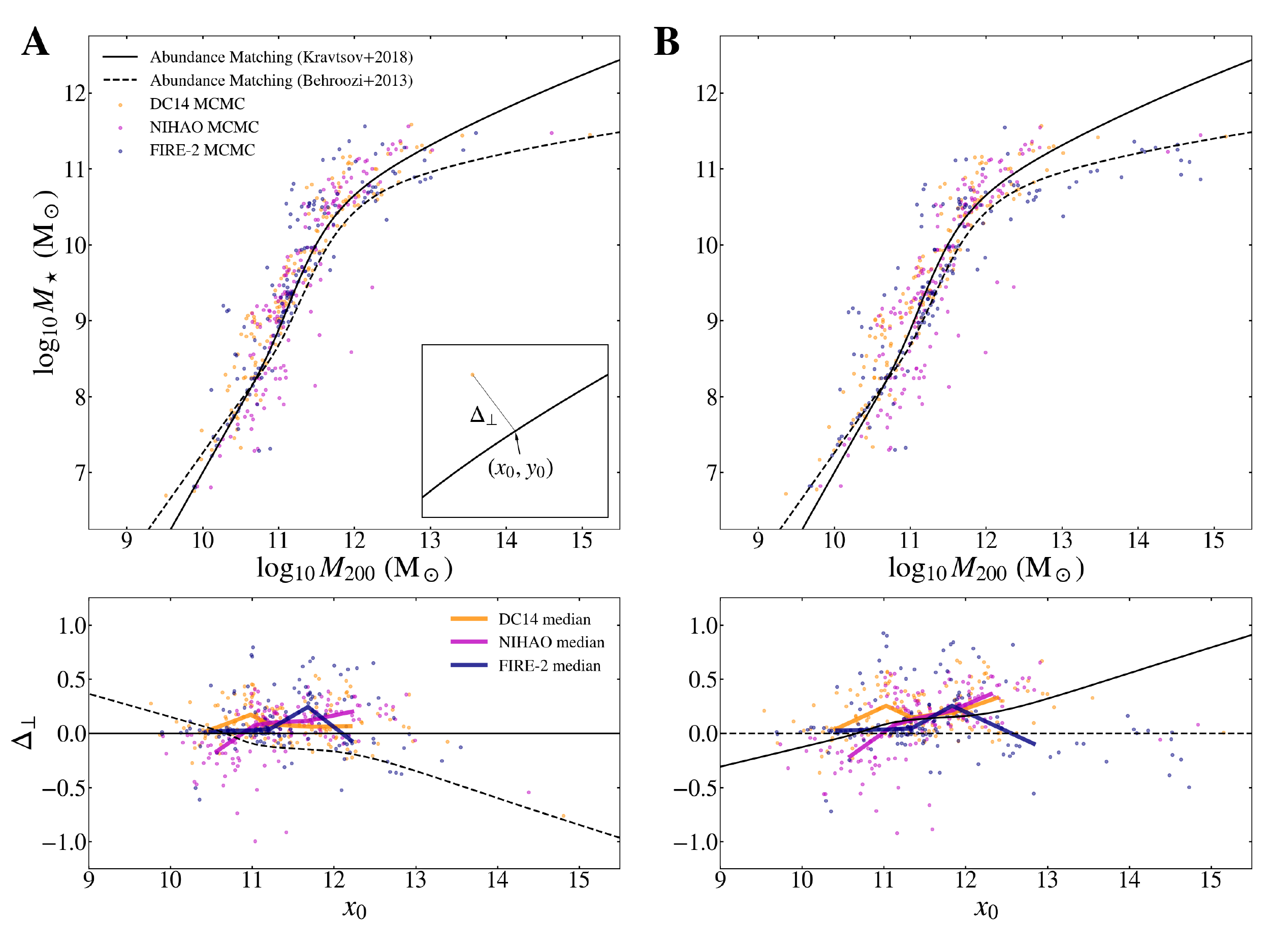}
    \vspace{-0.7truecm}
    \caption{\small 
    (A) Colored dots represent posterior values of stellar mass (${M_ \star }$) and halo mass (${M_{200}}$) from MCMC modeling results for three halo models. The prior relation between ${M_ \star }$ and ${M_{200}}$ is from the \cite{kravtsov2018} abundance matching relation. Orthogonal residuals shown in the bottom panel indicate that the posterior median relations agree with the prior relation. (B) The MCMC results based on the prior relation between ${M_ \star }$ and ${M_{200}}$ from \cite{behroozi2013}. Even in this case, the posterior median relations agree better with the \cite{kravtsov2018} abundance matching relation.
    } 
   \label{figS2}
\end{figure}

The stellar mass (${M_\star }$) – halo mass (${M_{200}}$)  relation is a major prior input in DM halo modeling. It is thus important to check whether posterior relations from MCMC results agree with the prior relation. Two prior ${M_\star } - {M_{200}}$ relations (that cover the range of currently available results in the literature) are used in MCMC modeling. Figure~\ref{figS2} shows the posterior relations from the MCMC results for three halo models. Because a scatter of 0.1 dex was allowed (for ${M_ \star }$ at fixed ${M_{200}}$) in MCMC modeling, the posterior values are scattered around the prior relation. When the prior from \cite{kravtsov2018} is used (Figure~\ref{figS2}(a)), the posterior results agree with the prior relation for all three models. Interestingly, even when the prior from \cite{behroozi2013} is used (Figure~\ref{figS2}(b)), the posterior results still better agree with the \cite{kravtsov2018} relation rather than the \cite{behroozi2013} relation. The SPARC galaxies prefer the \cite{kravtsov2018} relation.

\subsection{Halo Concentration}

When DM halos from DM-only simulations are described by the NFW profile (the case of $(\alpha ,\beta ,\gamma ) = (1,3,1)$ in Equation~(\ref{eq:zhao})) or the Einasto profile (the case of ${r_c} = 0$ in Equation~(\ref{eq:S15})), the scale radius ${r_s}$ or ${\tilde r_s}$ is the radius where the logarithmic slope is $ - 2$. The concentration ${c_{200}}( = {r_{200}}/{r_s})$ or ${\tilde c_{200}}( = {r_{200}}/{\tilde r_s})$ is well predicted from DM-only simulations under the $\Lambda {\rm{CDM}}$ cosmology. However, according to $\Lambda {\rm{CDM}}$ hydrodynamic simulations of galaxy formation, the baryonic physics of galaxy formation is expected to modify the halo mass distribution to a varying degree so that the profile is no longer NFW or Einasto in general. Consequently, in baryonic feedback-affected halos the scale radius ${r_s}$ or ${\tilde r_s}$ no longer corresponds to the radius where the slope is $ - 2$. For this reason, ${c_{200}}$ or ${\tilde c_{200}}$ is not constrained in MCMC modeling. However, an indirect constraint on ${c_{200}}$ was presented from the DC14 simulations as a scaling relation for ${c_{ - 2}}/{c_{{\rm{NFW}}}}$ given by Equation~(\ref{eq:S13}) where ${c_{ - 2}}$ is the radius where the slope is $ - 2$ for the modified halo mass profile. This constraint was used only for the DC14 modeling.

\begin{figure}
  \centering
  \includegraphics[width=0.7\linewidth]{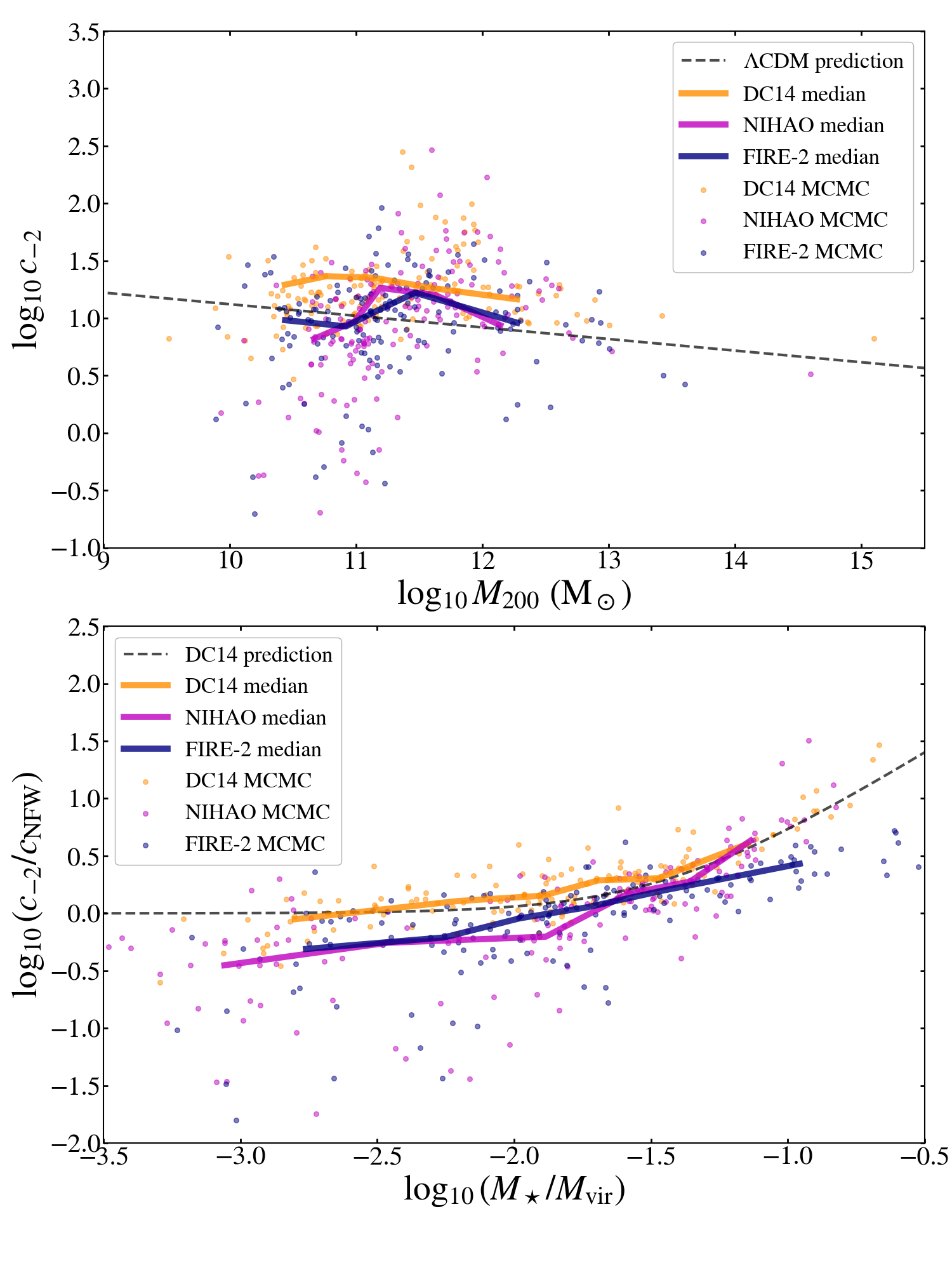}
    \vspace{-0.5truecm}
    \caption{\small 
    (Upper panel) Colored dots represent posterior values of ${c_{ - 2}}( \equiv {r_{200}}/{r_{ - 2}})$ where ${r_{ - 2}}$ is the radius where the logarithmic slope of the halo density is $ - 2$. The posterior values are compared with the NFW \citep{NFW} concentration from a $\Lambda {\rm{CDM}}$ DM-only simulation assuming the Planck cosmology \citep{dutton2014}. (Lower panel)  The posterior values of the ratio ${c_{ - 2}}/{c_{{\rm{NFW}}}}$ (where ${c_{{\rm{NFW}}}}$ is the NFW concentration from the $\Lambda {\rm{CDM}}$ DM-only simulation) are compared with the DC14 simulations prediction \citep{dicintio2014b} represented by the dashed curve. Note that the prior relation was imposed only in obtaining the DC14 MCMC results.
    } 
   \label{figS3}
\end{figure} 

Figure~\ref{figS3} shows the posterior values of ${c_{ - 2}}$ derived from the MCMC modeling results for the three halo models. The upper panel shows the posterior relation of ${c_{ - 2}}$ with ${M_{200}}$ and compares it with the prediction from the $\Lambda {\rm{CDM}}$ DM-only simulation \citep{dutton2014} under the Planck cosmology. The DC14 median is above the $\Lambda {\rm{CDM}}$ prediction for the whole range of ${M_{200}}$. The NIHAO and FIRE-2 medians are lower than the $\Lambda {\rm{CDM}}$ prediction for ${M_{200}} \mathbin{\lower.3ex\hbox{$\buildrel<\over
{\smash{\scriptstyle\sim}\vphantom{_x}}$}} {10^{11}}{M_\odot }$ but higher for ${M_{200}} \mathbin{\lower.3ex\hbox{$\buildrel>\over
{\smash{\scriptstyle\sim}\vphantom{_x}}$}} {10^{11}}{M_\odot }$. The lower panel of Figure~\ref{figS3} shows the posterior behavior of ${c_{ - 2}}/{c_{{\rm{NFW}}}}$ and compares it with the DC14 prediction. The DC14 MCMC posterior median agrees with or slightly exceeds the DC14 prediction, but the NIHAO and FIRE-2 posterior medians are lower than the DC14 prediction. 

\subsection{Density Slope between $r = 0.01{r_{200}}$ and $0.02{r_{200}}$}

\begin{figure}
  \centering
  \includegraphics[width=0.9\linewidth]{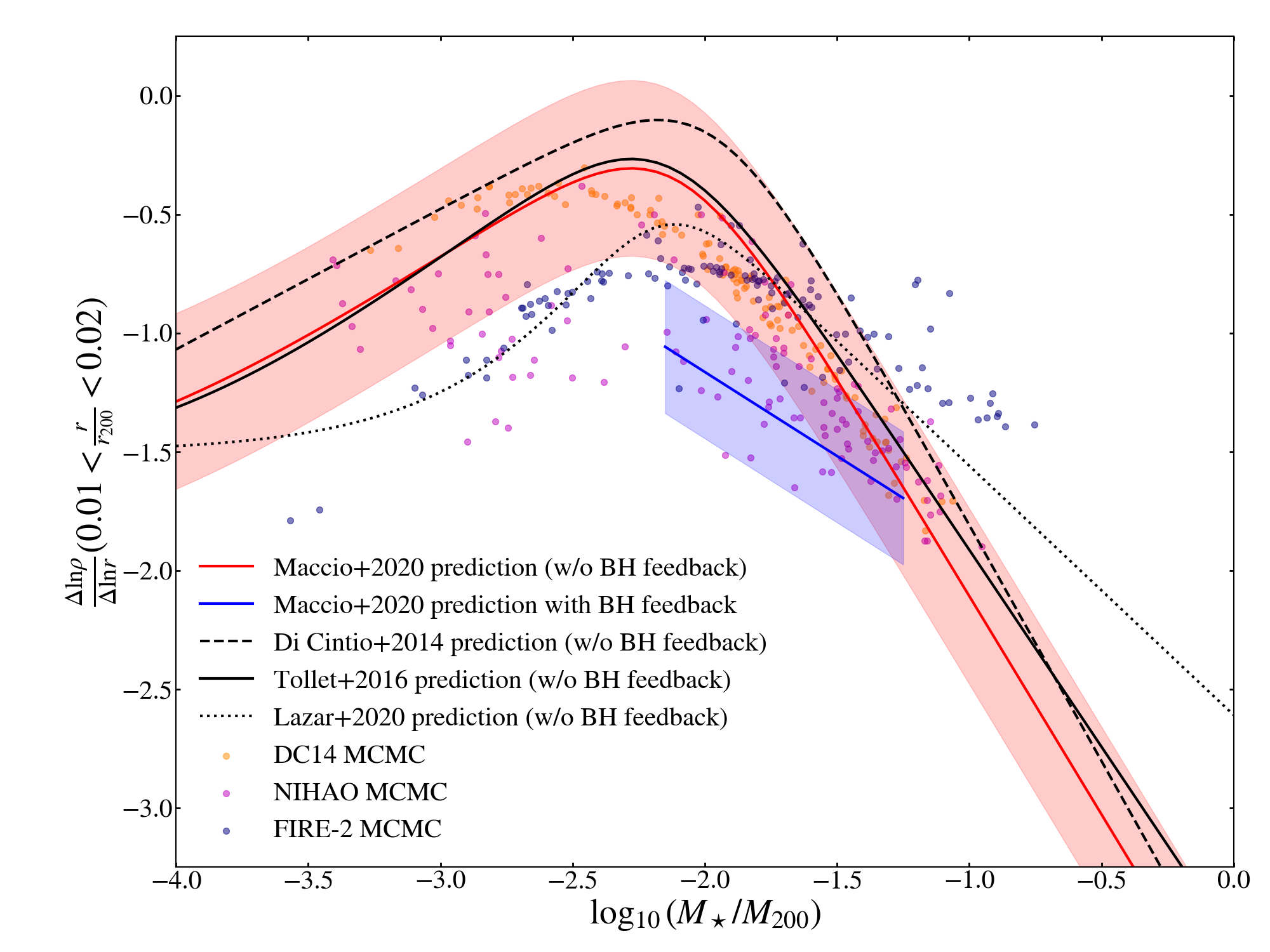}
    \vspace{-0.2truecm}
    \caption{\small 
    Posterior values of the logarithmic slope of the halo density between $r = 0.01{r_{200}}$ and $0.02{r_{200}}$ are compared with the prediction from various cosmological hydrodynamic simulations of galaxy formation. The red band represents the uncertainty from the \cite{maccio2020} simulations without black hole (BH) feedback. This band roughly covers the range from all previous simulations without BH feedback. The blue band represents the uncertainty for halos with ${M_{{\rm{200}}}} \mathbin{\lower.3ex\hbox{$\buildrel>\over {\smash{\scriptstyle\sim}\vphantom{_x}}$}} 4 \times {10^{12}}{M_\odot }$ from the \cite{maccio2020} simulations with BH feedback.
    } 
   \label{figS4}
\end{figure}

Various hydrodynamic simulations of galaxy formation in the literature \citep{dicintio2014a,tollet2016,lazar2020,maccio2020} provide values of the logarithmic slope of the mass density of the baryon feedback-affected halos. The values have large uncertainties though some generic pattern is seen in all simulations as shown in Fig.~\ref{figS4}.  Only for the NIHAO model was the predicted slope used as a prior for self-consistency. The more recent study by \cite{maccio2020} considered both cases with and without black hole (BH) feedback, while the earlier studies \citep{dicintio2014a,tollet2016,lazar2020} considered only the case without BH feedback. This can be important for massive galaxies with halo mass ${M_{200}} \mathbin{\lower.3ex\hbox{$\buildrel>\over
{\smash{\scriptstyle\sim}\vphantom{_x}}$}} 4 \times {10^{12}}{M_\odot }$, but the majority of SPARC galaxies are not so massive (see Figure~\ref{figS2}). The likely range from the \cite{maccio2020} case without BH feedback represented by the red band roughly includes all current results.

The DC14 MCMC results, although not constrained by any of the predicted slopes, reveal a well-defined relation between the slope and the ratio ${M_ \star }/{M_{200}}$. This relation is well within the red band from \cite{maccio2020}.  Note also that the DC14 results are in better agreement with the SPARC data than the NIHAO and FIRE-2 results as shown in Figure~\ref{RAR_DM}. The NIHAO and FIRE-2 modeling results give a somewhat scattered distribution of the slope. The NIHAO modeling results match better with the blue band (the case with the BH feedback). The FIRE-2 results roughly match the \cite{lazar2020} prediction based on the FIRE-2 simulations. This indicates self-consistency for the FIRE-2 results. 

\subsection{Scaling of Three Slope Parameters of the DC14 Model}

\begin{figure}
  \centering
  \includegraphics[width=0.6\linewidth]{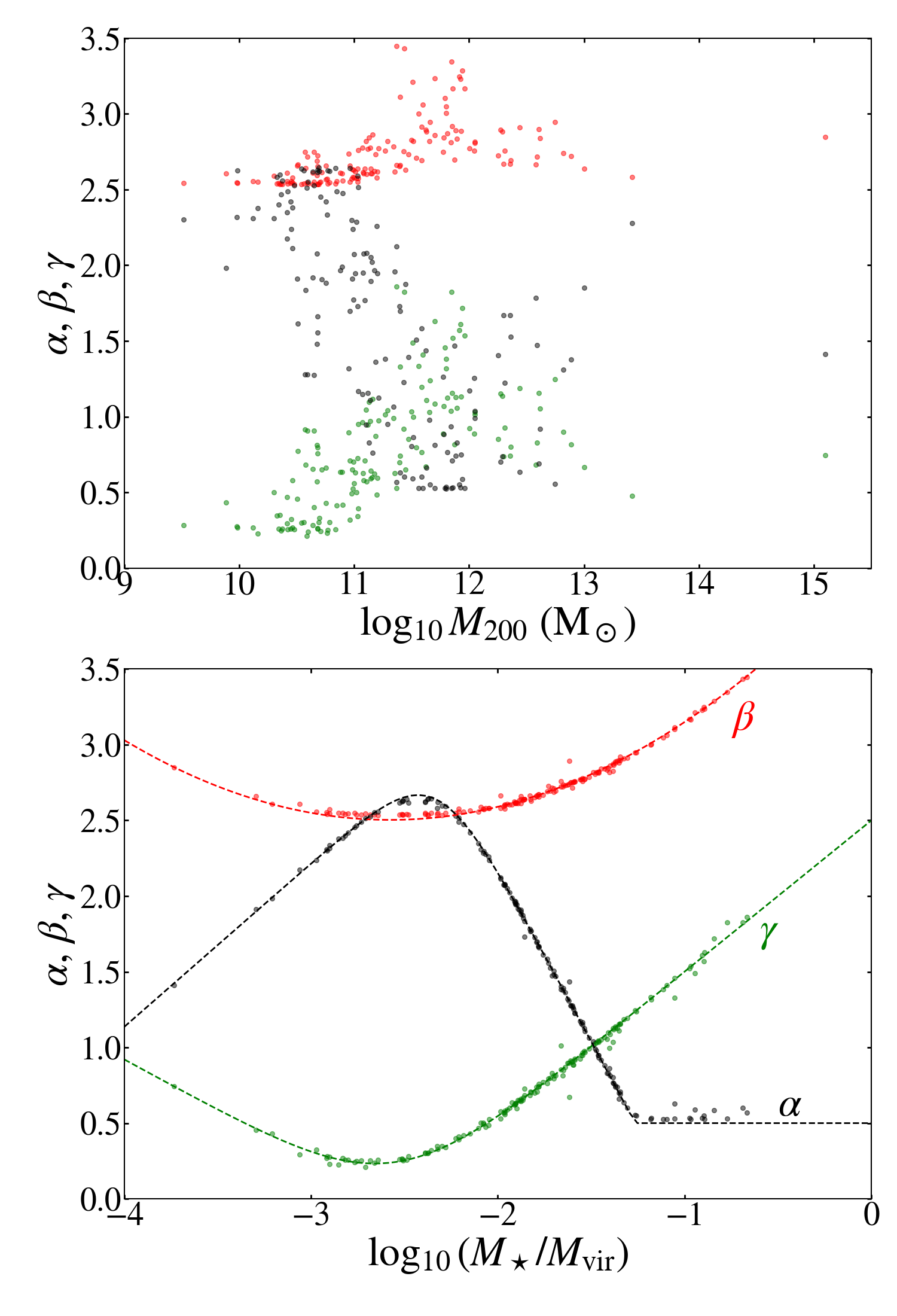}
    \vspace{-0.2truecm}
\caption{\small 
    Colored dots represent posterior values of three indices of the halo model given by Equation~(\ref{eq:zhao}) from the DC14 modeling results. The bottom panel compares the posterior values with the DC14 prior values \citep{dicintio2014b} represented by dashed curves.
    } 
   \label{figS5}
\end{figure}

The DC14 \citep{dicintio2014b} simulations provide scaling relations of three indices $\alpha $, $\beta $, and $\gamma $ as given in Equation~(\ref{eq:S12}). These relations were used as priors with a small scatter of 0.05 in producing MCMC results for the DC14 model. Here we use Figure~\ref{figS5} to check whether the posterior results show any systematic deviation from the prior relations. Overall, the posterior results match adequately the prior relations. The inner slope $\gamma $ does not show any sign of systematic deviation. The outer slope $\beta $ tends to be a little higher than the prior value for ${\log _{10}}({M_ \star }/{M_{200}}) <  - 2$. The transition index $\alpha $ tends to be higher than the imposed minimum value of 0.5 for high values of ${\log _{10}}({M_ \star }/{M_{200}})$.

\end{document}